\documentclass[fleqn,usenatbib]{mnras}

\usepackage{newtxtext,newtxmath}

\usepackage[T1]{fontenc}
\usepackage{ae,aecompl}

\usepackage{graphicx}	
\usepackage{amsmath}	
\usepackage{amssymb}	
\usepackage{gensymb}

\title[Beam modelling with single pulses]{Understanding the radio beam of PSR~J1136+1551 through its single pulses}

\author[Oswald et al.]{
Lucy Oswald,$^{1}$\thanks{E-mail: lucy.oswald@physics.ox.ac.uk (LSO)}
Aris Karastergiou,$^{1,2,3}$
Simon Johnston$^{4}$
\\
$^{1}$Department of Astrophysics, University of Oxford, Denys Wilkinson Building, Keble Road, Oxford OX1 3RH, UK\\
$^{2}$Department of Physics, University of the Western Cape, Cape Town 7535, South Africa\\
$^{3}$Department of Physics and Electronics, Rhodes University, PO Box 94, Grahamstown 6140, South Africa\\
$^{4}$CSIRO Astronomy and Space Science, Australia Telescope National Facility, PO~Box~76, Epping NSW~1710, Australia.
}

\date{Accepted 2019 July 27. Received 2019 July 26; in original form 2019 June 27}

\pubyear{2019}

\begin{document}
\label{firstpage}
\pagerange{\pageref{firstpage}--\pageref{lastpage}}
\maketitle

\begin{abstract}
The frequency widening of pulsar profiles is commonly attributed to lower frequencies being produced at greater heights above the surface of the pulsar; so-called radius-to-frequency mapping.
The observer's view of pulsar emission is a 1D cut through a 3D magnetosphere: we can only see that emission which points along our line of sight. 
However, by comparing the frequency evolution of many single pulses positioned at different phases, we can build up an understanding of the shape of the active emission region. 
We use single pulses observed with the Giant Metrewave Radio Telescope to investigate the emission region of PSR~J1136+1551 and test radius-to-frequency mapping.
Assuming that emission is produced tangential to the magnetic field lines and that each emission frequency corresponds to a single height, we simulate the single pulse profile evolution resulting from the canonical conal beam model and a fan beam model. 
Comparing the results of these simulations with the observations, we conclude that the emission region of PSR~J1136+1551 is better described by the fan beam model. 
The diversity of profile widening behaviour observed for the single pulses can be explained by orthogonally polarized modes propagating along differing frequency-dependent paths in the magnetosphere.
\end{abstract}



\section{Introduction}

\defcitealias{Gangadhara2004}{G04}
\defcitealias{Hassall2011}{H12}
\defcitealias{Noutsos2015}{N15}
\defcitealias{Dyks2015}{D15}

Half a century on from the discovery of pulsars by \cite{HEWISH1968a},
a full understanding of the nature of the radio emission mechanism
remains elusive. Efforts to increase our understanding have progressed
side-by-side with improvements to telescope sensitivity, as has been
summarized in \cite{Karastergiou2015}. 
Recently the greatest advancements
have been obtained through telescope receivers
which increase the bandwidth of observation. Examples include the
upgrade to the GMRT (Giant Metrewave Radio Telescope), and the
Ultra-WideBand Low (UWL) receiver on the Parkes 64m radio telescope \citep[][in preparation]{Hobbs2019}, both completed
in 2018. However, it is ever clearer from multi-frequency
observations that the canonical picture of pulsar emission is not only
incomplete, but now also insufficient. The complex frequency structure
of pulsar profiles varies to such an extent that it can no longer be
ignored over these observational bandwidths. Pulsar timing requires a measurement of the absolute rotational phase of the star on the profile, and frequency-dependent profiles make this difficult. 
It is possible up to a point to create frequency-dependent templates
for measuring the time of arrival
using empirical techniques, such as fits to the eigenvector
decomposition of a pulsar's frequency evolution, as done by
\cite{Pennucci2018}. 
However, this does not provide insight into the physical processes taking place in a pulsar magnetosphere.

A pulsar is a rotating, magnetized neutron star with a dipolar magnetic field, the axis of which is inclined with respect to the axis of rotation. Its spin period defines the radius of its light cylinder: the distance from the rotation axis at which the co-rotational speed is the speed of light. Magnetic field lines extending beyond the light cylinder are open: they cannot close as this would require the magnetospheric plasma to travel faster than the speed of light. The polar cap is a region on the pulsar surface about the magnetic axis, the edge of which is defined by the last open field lines. In the canonical description of a pulsar, radio emission is thought only to be generated by active field lines which have their origins in the polar cap, producing a beam of emission which co-rotates with the pulsar. The pulse profile observed arises from this radio beam traversing the line of sight.

When studying pulsar radio emission, the key observables are full 
polarization measurements of the pulsar
signal as a function of time. In order to relate these observables to
the conditions that generated them we must transform from our
one-dimensional view along the line of sight to the three-dimensional
emission region of the pulsar.  \cite{Lyne1988} combined the integrated profiles of multiple pulsars to determine the average 2D emission region for the pulsar population at a single frequency. Here we use multiple single pulses from one pulsar to investigate its emission region, and how it evolves with frequency, along our only line of sight trajectory.

Below about 1~GHz, pulsar profiles become wider as frequency decreases. Individual components in the profile move away from each other, in addition to becoming broader \citep[e.g.][]{Cordes1978}. In multi-component profiles, the outer components shift more with decreasing frequency than those closer to the centre of the profile \citep[e.g.][]{Mitra2002}.
The observed widening of pulse profiles is commonly
attributed to radius-to-frequency mapping (RFM): the idea that lower
frequencies are emitted at greater heights in the pulsar
magnetosphere, corresponding to a larger beam opening angle due to the
flaring of the magnetic field lines. An interpretation for
the causes of such emission height variation was incorporated into the
theory of the pulsar magnetosphere created by \cite{Ruderman1975}. In
their model, the frequency of emission is dependent on the plasma
frequency of the magnetosphere, which itself varies with the density
of the plasma, so that emission height $r$ is related to frequency $f$
by $r \propto f^{-\frac{2}{3}}$.

In order to be able to link frequency widening to emission height,
we must have an idea about the shape of the region of the magnetosphere responsible
for generating the emission. 
\cite{Gangadhara2004} (subsequently \citetalias{Gangadhara2004}) modelled pulsar emission as being produced
by a set of active magnetic field lines and propagating along a path
that lies tangent to the field lines at the emission point.
A common assumption made is that
the active field lines responsible for the radio emission have circular
symmetry about the magnetic axis. This results in a hollow cone of emission,
the radius of which increases with height due to the flaring field lines.
The hollow cone model, developed by Rankin and collaborators \citep[see for example][]{Rankin1983}, was proposed to explain the symmetry seen in
many pulsar profiles. It has been
extended to incorporate multiple concentric hollow cones and a
pencil-beam ``core'' emission along the magnetic axis, to encompass
the variety of profile morphologies observed.
An alternative is the fan beam
model, proposed originally by \cite{Michel1987} and then re-proposed
separately by both \cite{Dyks2010} and \cite{Wang2014}. It describes the
emission region as being composed of elongated streams that follow the
paths of a group of active field lines. \cite{Karastergiou2007} developed a model which describes the statistics of the population of pulsar profiles. This model describes the active region as being composed of groups of active field lines with their footprints forming patches randomly distributed within a ring. These field lines produce emission at multiple discrete heights along the same field lines, for a given frequency. 

\begin{figure}
    \includegraphics[width=\columnwidth]{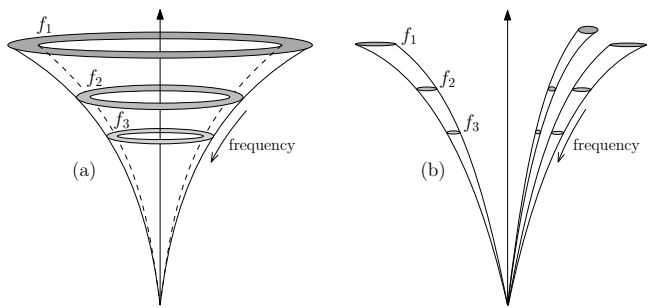}
    \caption{Two models of the pulsar radio emission region. For each diagram the vertical arrow represents the magnetic axis and the curved lines some region of active dipolar magnetic field lines. The shaded areas represent the parts of the emission region responsible for emission at discrete frequencies, labelled $f_{1}$, $f_{2}$ and $f_{3}$. Frequency of emission increases with decreasing height, as indicated by the labelled arrows. Diagram (a) shows the hollow cone model: a group of active field lines defining a ring on the pulsar surface. The fan beam model is shown in diagram (b): the active field lines form streams or fans.}
    \label{fig:models}
\end{figure}

Profile widening with frequency can be explained within these models by assuming a radius-to-frequency mapping such that lower frequencies are produced at greater heights. The simplest such relationship is to assume a one-to-one mapping, so that a single frequency is produced at a single height. For the hollow cone and fan beam models, this would result in the emission region for a particular frequency being ring-shaped and patchy respectively. The shapes and positions of the emitting regions will evolve with frequency, as is shown in Fig. \ref{fig:models}. 

More realistically, we might expect emission at a given frequency to be produced over a range of heights. \cite{Dyks2015}, hereinafter \citetalias{Dyks2015}, describe how this could work for the fan beam model such that RFM is still observed. 
\citetalias{Gangadhara2004} presents an alternative consideration, drawing on the work of \cite{Ruderman1975} to describe the curvature radiation produced by relativistic particles. In this model the characteristic frequency of curvature radiation is related to the Lorentz factor $\gamma$ of the emitting particle and the radius of curvature $\rho_{c}$ of the field line along which the particle travels:
\begin{equation}
    f = \frac{3c}{4\pi}\frac{\gamma^{3}}{\rho_{c}}.
\label{eq:f_rho}
\end{equation}
As \citetalias{Gangadhara2004} demonstrate, this means that, for a given frequency, emission produced closer to the magnetic axis must be produced at a lower height than that produced further from the magnetic axis.

An alternative, or possibly additional, piece of theory behind the
frequency widening of pulsar profiles is the
frequency-dependent refraction of orthogonally polarized plasma modes
(OPMs) in the magnetosphere. 
\cite{Melrose1977} and \cite{Melrose1979} proposed magnetospheric refraction as a mechanism for the spatial separation of OPMs.
Whereas the path of propagation of the X mode is unaffected by the plasma, the subluminous O mode is refracted (\citealt{Barnard1986,Weltevrede2003}). The subluminous O mode does not escape the magnetosphere due to Landau damping, however it can be converted to the superluminous O mode, which is refracted less and does escape. The transition radius where this conversion takes place is frequency dependent, meaning that lower frequencies are refracted more strongly than higher frequencies.
This increased path divergence at lower frequencies would therefore lead to a wider observed pulse profile. It may also explain the depolarization seen at higher frequencies: since the
OPMs are diffracted less they follow more similar paths
and cancel each other out in polarization.

Attempts to investigate the theory of RFM have always previously
focused on the frequency evolution of integrated pulsar
profiles. 
In particular, there is a
considerable history of studying PSR~J1136+1551, also known as PSR B1133+16. Recent 
examples include papers by \cite{Hassall2011} and \cite{Noutsos2015}
(hereinafter referred to as \citetalias{Hassall2011} and
\citetalias{Noutsos2015} respectively).
Their work focused
on trying to reconcile the different emission heights inferred through measuring 
both aberration/retardation and frequency widening. In the latter case
this was done by adopting the hollow cone model and
fixing the two peaks of the integrated profile to the last open field
lines. 

Emission heights calculated using RFM are
dependent on the
model used to describe the emission region. Since the integrated
profile is in effect a statistical description of the pulsar emission,
averaging many single pulses to produce a stable profile, it gives us
limited capacity to test the validity of the model assumed. However,
using the single pulses of PSR~J1136+1551 gives us a population of
instances where we can trace the pulsar emission along the magnetic
field lines, allowing us to compare different emission models of the
pulsar magnetosphere through a statistical approach. 

We simulate single pulses that are generated in active regions in the magnetosphere
that evolve with frequency. We can then compare the frequency evolution resulting from
our emission region model to that we observe.
To do this,
we reduce both our observed and simulated single pulses to a set of metrics that are 
easy to measure and understand within the context of the assumptions clearly laid out in Section \ref{sec:method}.
Of key importance in our simulations is the fact that the emission region model
allows us to use each single pulse at a reference
frequency to predict its location at other frequencies. The first
metric is therefore the distribution of the locations of the subpulses
for all single pulses at all frequencies, for observed and simulated
data, captured through the means and standard deviations of these
distributions. In the context of RFM, the second metric is
related to the separation of subpulses at different frequencies,
defined as the pulse phase separation of the subpulse centroids for
clearly double-peaked pulses. As we describe in section \ref{sec:method}, we capture the
frequency dependence of pulse separation by means of a power law spectral index. The
simulations with different emission region models lead to clear predictions of the spectral index
distribution, as well as its dependence on subpulse separation, both
of which we use as metrics to evaluate the tested models.

Our observations of the single pulses and the data analysis techniques
we employed are described in sections \ref{sec:obs} and
\ref{sec:analysis}. We built simulations of pulsar magnetosphere
emission, encompassing the hollow cone and fan beam models and the
frequency evolution of OPMs: the details are
given in section \ref{sec:sim}. Section \ref{sec:results} presents the
comparisons made between the data and the simulations, the
ramifications of which are discussed in section \ref{sec:disc}. We
give our conclusions in section \ref{sec:conc}.

\section{Observations}
\label{sec:obs}

The GMRT \citep{Swarup1991}
is an aperture-synthesis interferometric
array of consisting of thirty 45~m antennas, spread over a 25~km region
80~km north of Pune, India.  The Y-shaped array is composed of a
central core of 14 antennas within 1 km$^2$ with the rest spread along
the arms.  The array can also be used as a single-dish telescope in
the so-called phased-array configuration, where signals from
various antennas can be added coherently (in phase).  For our
purposes we used the phased-array mode, and to minimize the effect of spatial
variation of the ionosphere we used the central core and first two
antennas in each of the arms to make up the phased-array. The GMRT operates at
low radio frequencies and has been recently upgraded \citep[see][]{Gupta2017}
to have four broad bands namely Band-5: 1050--1450 MHz, Band-4: 550--850 MHz,
Band-3: 250--500 MHz, and Band-2: 120--250 MHz. We used Band-3, where the
feeds are linearly polarized and are converted to left and right
circulars using hybrids. Subsequently, per antenna, the signals are
amplified and the left and right circular channels are further
downconverted into baseband signals, after which the voltages are
sampled at the Nyquist rate and fed to the GMRT Wideband Backend (GWB).  
We recorded data in the full-polar mode,
where currently the maximum available bandwidth is 200~MHz, and we used
the frequency range 300--500~MHz. The digitized Nyquist sampled signals are then
Fourier transformed to obtain 2048 spectral channels per polarization
and the selected set of antennas are added 
in phase by the GWB beamformer. In the full Stokes phased-array mode the 
beamformer computes the auto and cross polarized power and 
can also average the signals to reduce the output data rate. We recorded full Stokes data at a time resolution of 327.68 $\mu$s. 

Our observing strategy was as follows. We initially observed the
flux calibrator 3C286 which was also used for initial phasing of the array. 
Then before observing the pulsar PSR~J1136+1551 we
checked the validity of the phasing using the nearby phase calibrator 
3C241, and if needed phasing was redone. Uncalibrated raw data in 
filterbank format consisting
of auto and cross products were recorded on the pulsar.

The initial processing of the data was done using the software package
{\sc dspsr} developed by \cite{VanStraten2011}, to generate incoherently
dedispersed single pulses which could be analysed using the software
package {\sc psrchive} \citep{Hotan2004}. The data products, four coherence
parameters, were gain calibrated using software written for this
purpose and converted into Stokes parameters. For the purposes of this
work we used only the total intensity, Stokes I. We performed the
dedispersion with a DM of 4.926~cm$^{-3}$pc,
and binned the data into 3600~bins across the pulse period. This gives a time
resolution of 0.3~ms and a phase resolution of 0.1$\degr$. Finally, we reduced the data into ten
frequency channels of 20~MHz each across the observing band of
300--500~MHz. We discarded three of these channels due to RFI---those
at 310, 370 and 490~MHz---leaving seven channels remaining. The data
are plotted as a function of phase $\phi$, centred at $\phi=0\degree$.

Many of the single pulses are contaminated by an oscillating signal
with a frequency of 50~Hz, suggesting the presence of a mains
source. It was found not to be possible to remove the signal from
the data, since the peaks in the single pulse profile result in a peak
in Fourier space that is too close to the 50~Hz peak. The magnitude of
this signal fluctuates with time, so that some pulses
are strongly affected whilst others are hardly affected at all. Those
pulses strongly contaminated by the 50~Hz signal were excluded from
the analysis.

\section{Data analysis}
\label{sec:analysis}

\subsection{Double component single pulses}
\label{sec:double}

\begin{figure}
    \includegraphics[width=\columnwidth]{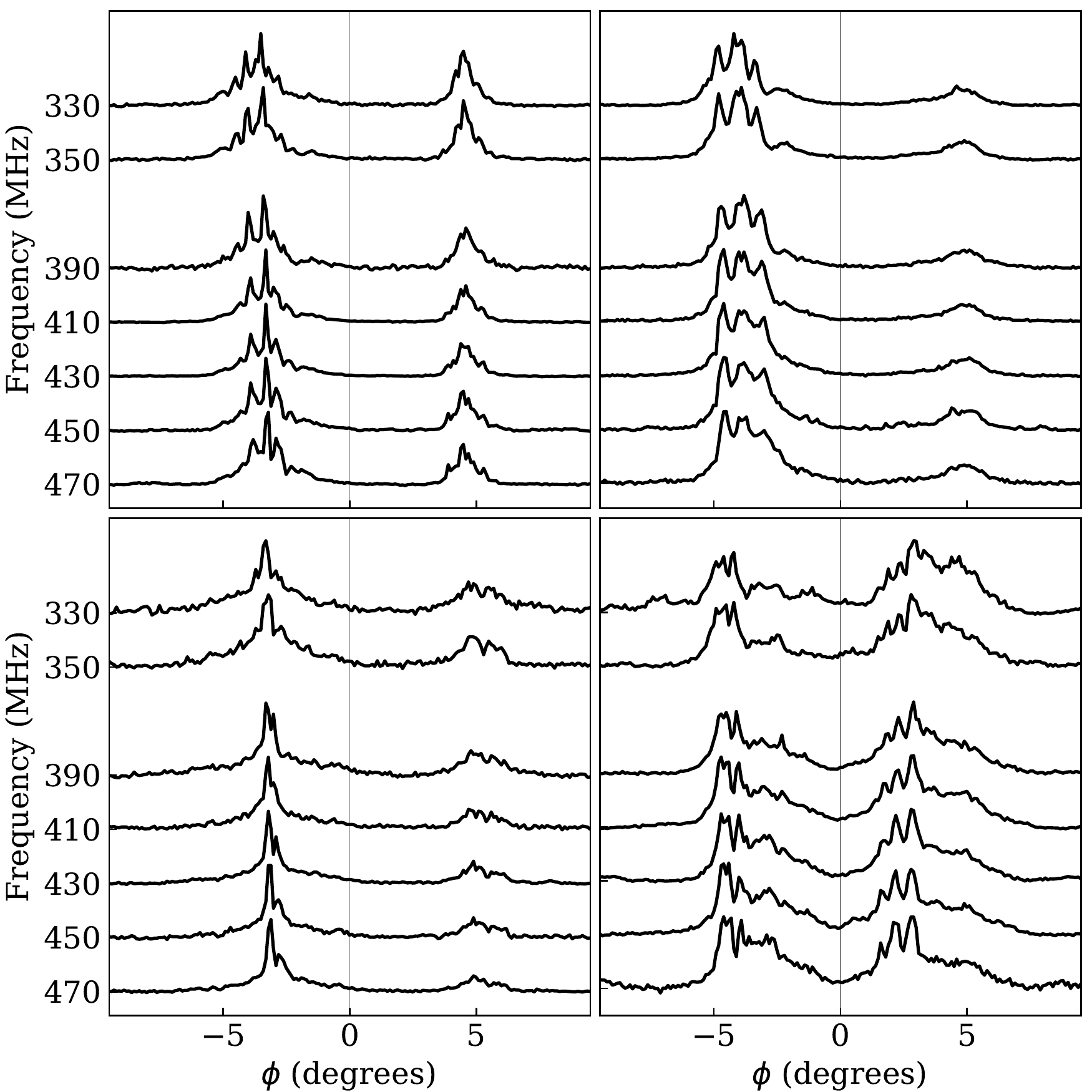}
    \caption{Profiles of four example single pulses used in the analysis with the seven frequency channels plotted vertically with the highest frequency at the bottom. The pulses have been dedispersed and aligned with respect to the defined fiducial plane (at phase $\phi = 0\degr$) as described in the text.}
    \label{fig:singlepulses}
\end{figure}

In this research we look to characterize the frequency widening of single pulses through
simulating how they evolve with frequency and comparing this to the observational results.
We therefore used only those pulses that were clearly double-peaked, since it is
only for these that we can clearly identify the
intrinsic frequency widening through measuring the changing peak separation. Some example single pulses are shown in Fig. \ref{fig:singlepulses}.

We identified the edges of the on-pulse region of the
integrated profile as follows. We smoothed the normalized integrated
profile with a median filter and
then calculated the signal to noise ratio (SNR), assuming Gaussian noise, by subtracting the
median value of the profile and dividing by $\sigma = 1.4826*$MAD where MAD is the
median absolute deviation. The profile edges were then
deemed to be the first and last bins where SNR > 4, a limit chosen by eye so
that the edges were located in the regions where the smoothed flux
gradient began to change. The width of any given
single pulse is smaller than the width of the integrated
profile, so by using the edges of the integrated profile to define the
on-pulse region we can be sure that we do not cut out flux
from the single pulses.

\begin{figure*}
    \includegraphics[width=\textwidth]{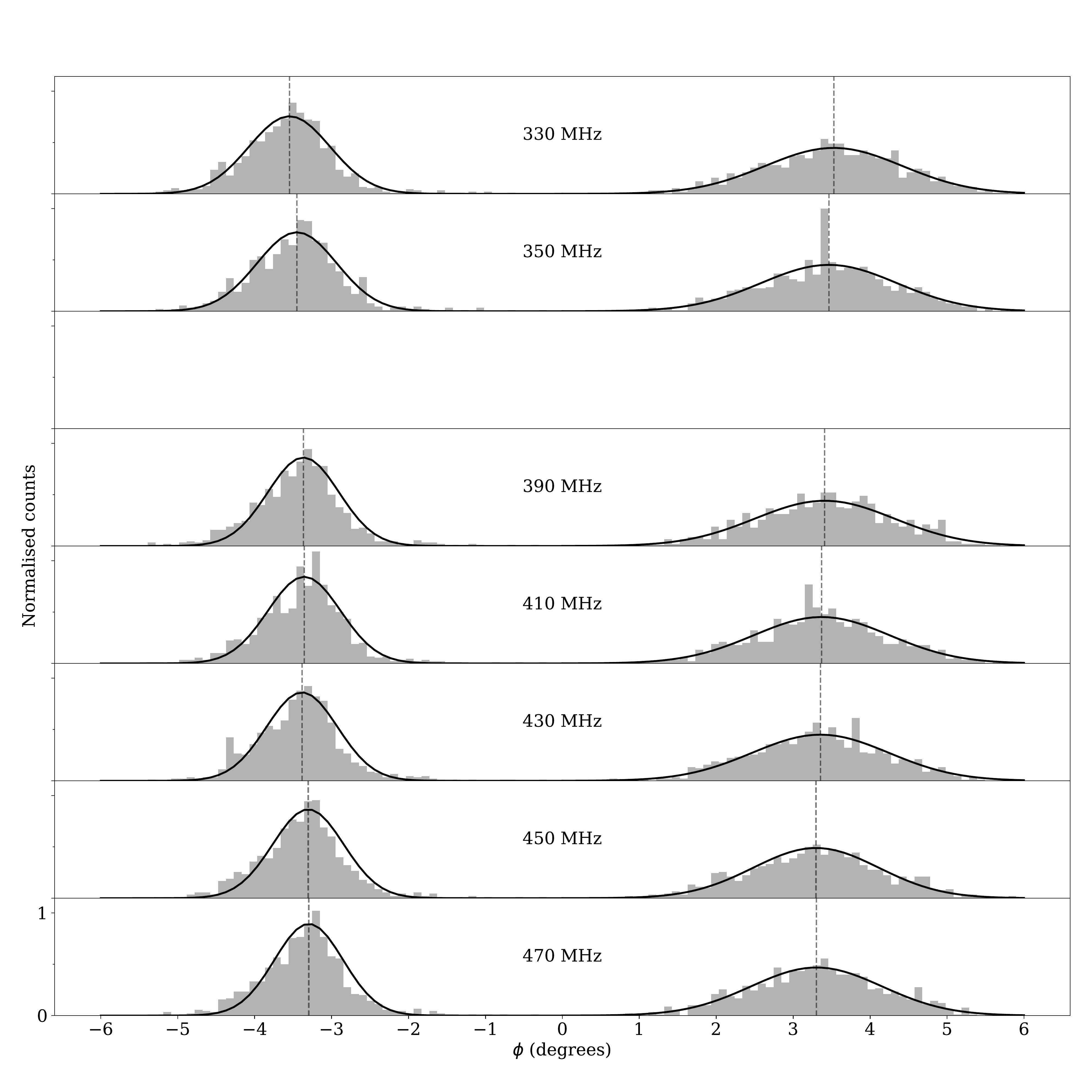}
    \caption{Histograms of the positions of the subpulses across the seven frequency bands observed. Vertical lines mark the means of the distributions to guide the eye to the distribution broadening. A space is left to mark the frequency band removed due to excess noise, so that the histograms are evenly spaced with frequency along the y axis. Gaussian fits to the subpulse distributions are over-plotted as black lines. The means and standard deviations of these Gaussians are given in Table \ref{tab:musig}.}
    \label{fig:datahist}
\end{figure*}

We correlated the on-pulse region of each pulse with a Gaussian located exactly half-way between the edges of the on-pulse region, with a standard deviation equal to that of a Gaussian fit to the widest peak of the integrated profile, a value of $\sigma = 1.53~\degr$,
and a normalized amplitude of $1/\sqrt{2 \pi \sigma ^{2}}$. 
The correlation smooths the pulse profile, allowing us to easily identify the positions of the subpulse peaks as turning points in the correlation function. Accordingly, any pulse that had exactly two peaks in the correlation function with a magnitude greater than our chosen cut-off of 0.1 
in every frequency channel, was defined as clearly double-peaked. We estimated the error on the peak positions as the distance between the peak location and the position with the greatest flux in each half of the profile separately. 
Pulses contaminated with RFI were discarded.
Of the original 4,759 pulses recorded, this left 885 pulses for the RFM analysis. 
Fig. \ref{fig:datahist} shows histograms of the positions of the subpulse peaks at each frequency, with Gaussian fits to the distributions overlaid. The parameters describing these Gaussians are given in Table \ref{tab:musig}. It is clear that a net divergence and broadening of the subpulse distributions is seen as we move to lower frequencies.

\subsection{Fiducial plane}
\label{sec:fid}

Taking only the double-peaked pulses and plotting histograms of the positions of the subpulses at each frequency, 
we realized that the mid-point between the two subpulse distributions was shifted successively earlier in time at lower frequencies.
We applied an adjustment to the DM of 0.034 cm$^{-3}$pc, giving a total DM of 4.892 cm$^{-3}$pc.
Once this was applied, the two distributions of subpulses moved symmetrically outwards with decreasing frequency away from their mid-point. This may be interpreted as the subpulses being produced from two emission regions positioned equidistant from the magnetic axis. It is common practice to place the fiducial plane halfway between the peaks of the integrated profile of PSR~J1136+1551, on account of its near-symmetry.

The results we draw from the data depend on the separation of the subpulses in each single pulse and are therefore independent of our choice of DM and positioning of the fiducial plane. The simulation however requires this information as a starting point for fixing the subpulses to field lines within the pulsar magnetosphere.

\section{Simulations}
\label{sec:sim}

\subsection{Method}
\label{sec:method}

In order to model the pulsar radio emission, we make the following assumptions:
\begin{itemize}
    \item The pulsar magnetosphere has a \textbf{dipolar} magnetic field.
    \item Emission is produced \textbf{tangential} to open magnetic field lines.
    \item Each frequency of emission is produced at a \textbf{single height}.
    \item Lower frequencies originate from higher heights: this is \textbf{Radius-to-Frequency Mapping} (RFM).
\end{itemize}
Given these assumptions, a set of active field lines is required to produce profiles with components of finite width.
We identify the field lines in terms of a constant, $K$, taken from: 
\begin{equation}
    r = K\sin^{2}\theta.
    \label{eq:r}
\end{equation}
This gives us an additional assumption:
\begin{itemize}
    \item The emission is generated by some \textbf{region of active field lines}. We define this as either a ring centred on the magnetic axis, or one or more circular patches.
\end{itemize}

\cite{Karastergiou2001} computed a bin-by-bin cross-correlation function between the single pulse profiles of PSR B0329+54 at 1.41 and 2.69~GHz. The strong correlation showed that subpulses at different frequencies are likely to have been produced by the same mechanism. The subpulses of PSR J1136+1551 similarly appear broadband, which leads to our final assumption:
\begin{itemize}
    \item The \textbf{same} active field lines at different heights are responsible for a subpulse observed across a broad band.
\end{itemize}

The 3D emission region is then a combination of this 2D shape with the frequency variation along the field line, resulting in either the hollow cone model or the fan beam model. Fig. \ref{fig:models} compares visualizations of the emission regions for these two cases.
An alternative would be to replace the constant emission height assumption with an assumption that emission frequency is related to field line curvature, as in equation \ref{eq:f_rho}. Such a picture is discussed by \citetalias{Gangadhara2004}, but we do not address it here.

At a given frequency a subpulse is produced by a group of active field lines at a given height. The shape of the observed subpulse depends on how the line of sight cuts this region. It is therefore important to distinguish between the observed position of the peak along the line of sight, $p_{\rm obs}$, and the position of the peak of the emission region, $p_{\rm peak}$, as there is no certainty that these correspond. 

$p_{\rm obs}$ and $p_{\rm peak}$ refer to the positions of these peaks within the 2D projection of the pulsar magnetosphere perpendicular to the magnetic axis, as visualized in figures \ref{fig:hc_projection} and \ref{fig:fan_projection}. By nature of being observable, $p_{\rm obs}$ lies upon the line of sight, whereas $p_{\rm peak}$ marks the absolute peak of emission and therefore may be coincident with $p_{\rm obs}$, or may not lie upon the line of sight.

Our final assumption means that at different frequencies the position of the emission peak will be shifted along the active field line responsible for its production. The position of the observed peak, $p_{\rm obs}$, will therefore also be shifted along the line of sight by an amount that depends on the relationship between $p_{\rm peak}$ and $p_{\rm obs}$. This relationship depends on the shape of the emission region. We then measure a corresponding phase shift in the position of the subpulse peak: this is the cause of the observed frequency widening.

For a single pulse at some reference frequency $f_{\rm ref}$, determining the active field lines then allows us to simulate the same pulse at all frequencies, based on the assumptions above and our choice of emission region model. This allows us to compare our two different models of the emission region shape---the hollow cone and fan beam models---to see which replicates the data more accurately.

Let us call the field line responsible for the peak of the overall emission region $F_{\rm peak}$. 
The steps of the modelling process are as follows. The equations used to convert between parameters are given in section \ref{sec:maths}.
\begin{enumerate}
    \item Choose the reference frequency, $f_{\rm ref} = 330$ MHz, the lowest frequency channel of our observing band.
    \item Identify $F_{\rm peak}$ in terms of its footprint $s^{p}_{L}$ upon the neutron star surface. We choose $s^{p}_{L} = 0.5$, an average of the values calculated for seven pulsars in \cite{Gangadhara2001} and \cite{Gupta2003}.
    \item Define $p_{\rm peak}$ = E($p_{\rm obs})$, the mean position of the distribution of observed subpulse peaks, at $f_{\rm ref}$. In doing so we are setting the peak of the emission region to be on the line of sight at the reference frequency. This means we can identify the emission height at this frequency using the mean phase $\mu$.
    \item Make an assumption about the shape of the emission region, so that we can relate $\mu$ to $p_{\rm peak}$ at all other frequencies.
    \item At every other frequency, calculate the emission height resulting from the observed phase of $\mu$.
    \item Use the observed phase of the subpulse peak in each single pulse at $f_{\rm ref}$ with the emission height at this frequency to calculate the footprint $s^{i}_{L}$ of the corresponding active field line. This requires the same assumption that the observed and emitted subpulse peak positions are the same at $f_{\rm ref}$.
    \item Use footprints $s^{i}_{L}$ and the emission heights to identify the predicted phases of the subpulse peaks at all other frequencies.
    \item Add a small Gaussian noise contribution to each subpulse position, with $\sigma = 0.1 \degr$.
    \item Measure the phase separation between the two subpulses of each single pulse at all frequencies and compare the simulation results to those of the data.
\end{enumerate}

We define $\eta$ as the spectral index associated with the frequency widening of subpulse separation, and we compute it by fitting a power law to the peak separation $sep$ against frequency $f$: $sep \propto f^{\eta}$. We also tried adding a constant term $sep \propto f^{\eta} + C$, as was done by \cite{Thorsett1991}, however this resulted in less good fits. Collectively we calculate the mean separation of the two subpulse distributions and the standard deviations of the distributions, $\sigma_{1}$ and $\sigma_{2}$, at each frequency, through fitting Gaussians to the histograms of subpulse positions. 

\subsection{The mathematics of the modelling process}
\label{sec:maths}

The mean phase, $\mu$, of each subpulse distribution can be related to the beam half-opening angle $\rho$ through spherical geometry as given in \cite{Gangadhara2001}:
\begin{equation}
    \cos{\rho} = \cos{\alpha}\cos{\left(\alpha+\beta\right)} + \sin{\alpha}\sin{\left(\alpha+\beta\right)}\cos{\left(\mu\right)}
    \label{eq:rho}
\end{equation}
where $\alpha$ is the angle between the magnetic and rotation axes and $\beta$ is the impact parameter: the angle between the magnetic axis and the line of sight at the fiducial plane. 
We use the values $\alpha = 51.3 \degree$ and $\beta = 3.7 \degree$, as derived by \cite{Lyne1988} through a fit of the rotating vector model (RVM) to the polarization position angle (PA) profile. 

We are aware of the shortcomings of RVM fits, which have a high degree of degeneracy between $\alpha$ and $\beta$ \citep{Rookyard2015}. 
The results presented in this paper are consistent under variation of $\alpha$ and $\beta$ by small amounts. Our models do suggest that a radically different geometry, such that the value of $\alpha$ is decreased considerably, would not be equally easy to accommodate. However, reconciling the narrowness of the pulse profile with a much decreased $\alpha$ also poses difficulties, and supports our choice here.
Future investigation, using polarization data, will reveal the extent to which the choice of emission model can provide constraints on the geometry, additional to fitting the RVM to the PA.

Half-opening angle $\rho$ is then converted to spherical polar angle $\theta$ through the following equation, given by \cite{Gangadhara2004}: 
\begin{equation}
    \cos(2\theta) = \frac{1}{3}\left(\cos\rho\sqrt{8 + \cos^{2}\rho} - \sin^{2}\rho\right), -\pi \leq \rho \leq \pi.
    \label{eq:theta}
\end{equation}
This equation is mathematically valid for the full range of $-\pi \leq \rho \leq \pi$, however, since beam half-opening angle is an absolute value, the value of $\rho$ obtained from equation \ref{eq:rho} is defined to lie within the range $0 \leq \rho \leq \pi$.

We use the superscript $j$ to refer to different frequencies. 
From the polar angle of the distribution mean at each frequency, $\theta_{\mu}^{j}$, we use our model of the emission region shape to obtain the polar angle of the peak of the emission region, $\theta_{\rm peak}^{j}$. Details of the conversion between $\theta_{\mu}^{j}$ and $\theta_{\rm peak}^{j}$ for each emission region model are given in sections \ref{sec:hcone} and \ref{sec:fan}. We then calculate the emission heights $r^{j}$ using equation \ref{eq:r},
where our reference field line $F_{\rm peak}$ has field line constant $K^{p}$. 

$K^{i}$, the constant identifying the $i^{th}$ field line of the emission, is described in terms of the footprint parameter ratio $s^{i}_{L}$ described in \cite{Gangadhara2001}. $s^{i}_{L}$ is the distance to the $i^{th}$ field line across the surface of the pulsar from the magnetic axis, $s^{i}$, divided by the equivalent distance to the last open field line, associated with the radius of the light cylinder, $s_{L}$. $K^{p}$ is given by 
\begin{equation}
    K^{p} = \frac{r_{s}}{\sin^{2}\left(s^{p}_{L}\arcsin{\sqrt{r_{s}/R_{LC}}}\right)}
    \label{eq:K}
\end{equation}
where $r_{s}$ is the radius of the neutron star, conventionally defined as 10 km, and $R_{LC} = \frac{cP}{2\pi}$ is the radius of the light cylinder, which for PSR~J1136+1551 is $\sim$57000 km. The full derivation of equation \ref{eq:K} is given in Appendix \ref{sec:appendix1}.

To calculate the footprints of all the subpulse peaks we use equations \ref{eq:r}, \ref{eq:rho}, \ref{eq:theta} and \ref{eq:K} as before, but now using $r$ to find $K$, rather than the other way around. 
Using equation \ref{eq:r} again, with our calculated emission heights and field line constants ($K^{i}$), we can now obtain the polar angles $\theta^{ij}$ for all subpulse peaks at all frequencies. We convert our emission polar angles back to the corresponding polar angles for the points that lie along the line of sight using our chosen model for the shape of the emission region (see sections \ref{sec:hcone} and \ref{sec:fan}). Inverting equations \ref{eq:rho} and \ref{eq:theta} we obtain the simulated subpulse phases across the frequency band, which we combine to make double-peaked pulses. 

\subsection{The hollow cone model}
\label{sec:hcone}

\begin{figure*}
    \includegraphics[width=\textwidth]{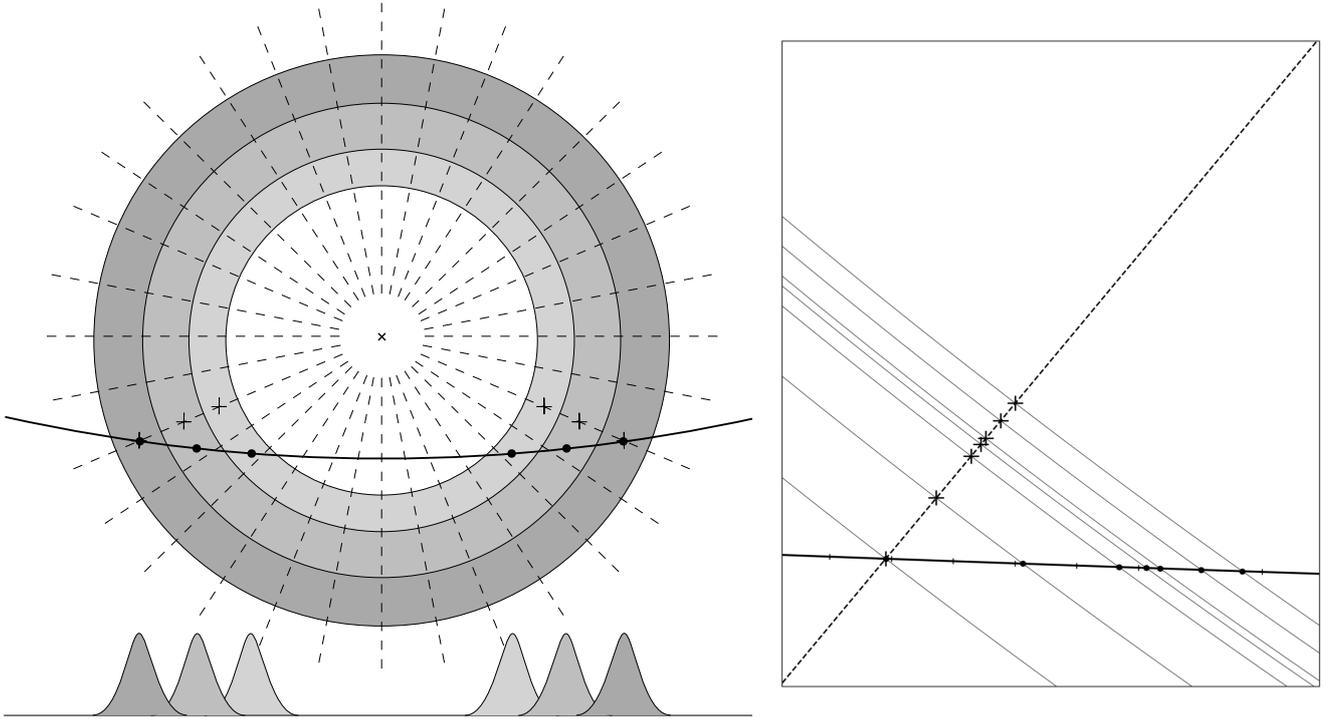}
    \caption{The figure on the left depicts the emission region in the hollow cone model, viewed as a projection down the magnetic axis of the pulsar, with the line of sight (solid) cutting selected field lines (dashed). The shaded regions indicate the projected emission region at three different frequencies, where lower frequency corresponds to darker shading and a larger ring diameter. Below the projection are the observed profiles that would result from the line of sight cutting the three different emission regions. 
    The positions where the line of sight cuts the peak of the ring emission are marked with points. Crosses mark the positions where the reference field line cuts the peak of the ring emission. 
    On the right a zoomed section of the line of sight (solid thick line) is displayed, with the reference field line $F_{\rm peak}$ (dashed). The mean observed positions of the subpulse distributions at all seven frequency channels are marked along the line of sight by black points. The inferred emission points along the reference field line are marked by crosses. They are connected to the observed mean positions by the circles of constant polar angle $\theta$ (solid thin lines). Tick marks along the line of sight (solid thick line) indicate intervals of $\frac{\pi}{4000}$ radians, going from $\frac{-73}{4000}\pi$ on the right to $\frac{-80}{4000}\pi$ on the left.}
  \label{fig:hc_projection}
\end{figure*}

The shape of the emission region gives us the relationship between the observed peak of the subpulse profile and the actual peak of the emission region. In the hollow cone model we assume that emission at a given frequency comes from a ring of field lines that is symmetrical about the magnetic axis. The field line responsible for the peak of the emission, $F_{\rm peak}$, is one of a ring of identical field lines producing identical emission. This means that the field lines responsible for the actual and observed emission peaks at a given frequency have the same field line constant $K$. From equation \ref{eq:r}, the same values of $r$ and $K$ result in the same spherical polar angle $\theta$: 
\begin{equation}
\theta_{\rm peak}^{j} = \theta_{\mu}^{j}.
\label{eq:hconetheta}
\end{equation}
Fig. \ref{fig:hc_projection} is a view down the magnetic axis, showing the 2D projections of the magnetic field lines and the path of the line of sight across the projection. It shows how the observed positions of the means of the subpulse distributions at all frequencies (points) are related to the emission points along the reference field line (crosses) through circles of constant spherical polar angle $\theta$. 

Modelling the emission region as a ring results in a symmetrical double-peaked profile, as the line of sight cuts the ring in two places. However, it is clear from Fig. \ref{fig:datahist} that the two subpulse distributions making up the single pulse profiles are of different shapes. We therefore relax the requirement that a ring of emission be absolutely symmetrical, and model the two subpulse distributions independently.

\subsection{The fan beam model}
\label{sec:fan}

\begin{figure*}
  \includegraphics[width=\textwidth]{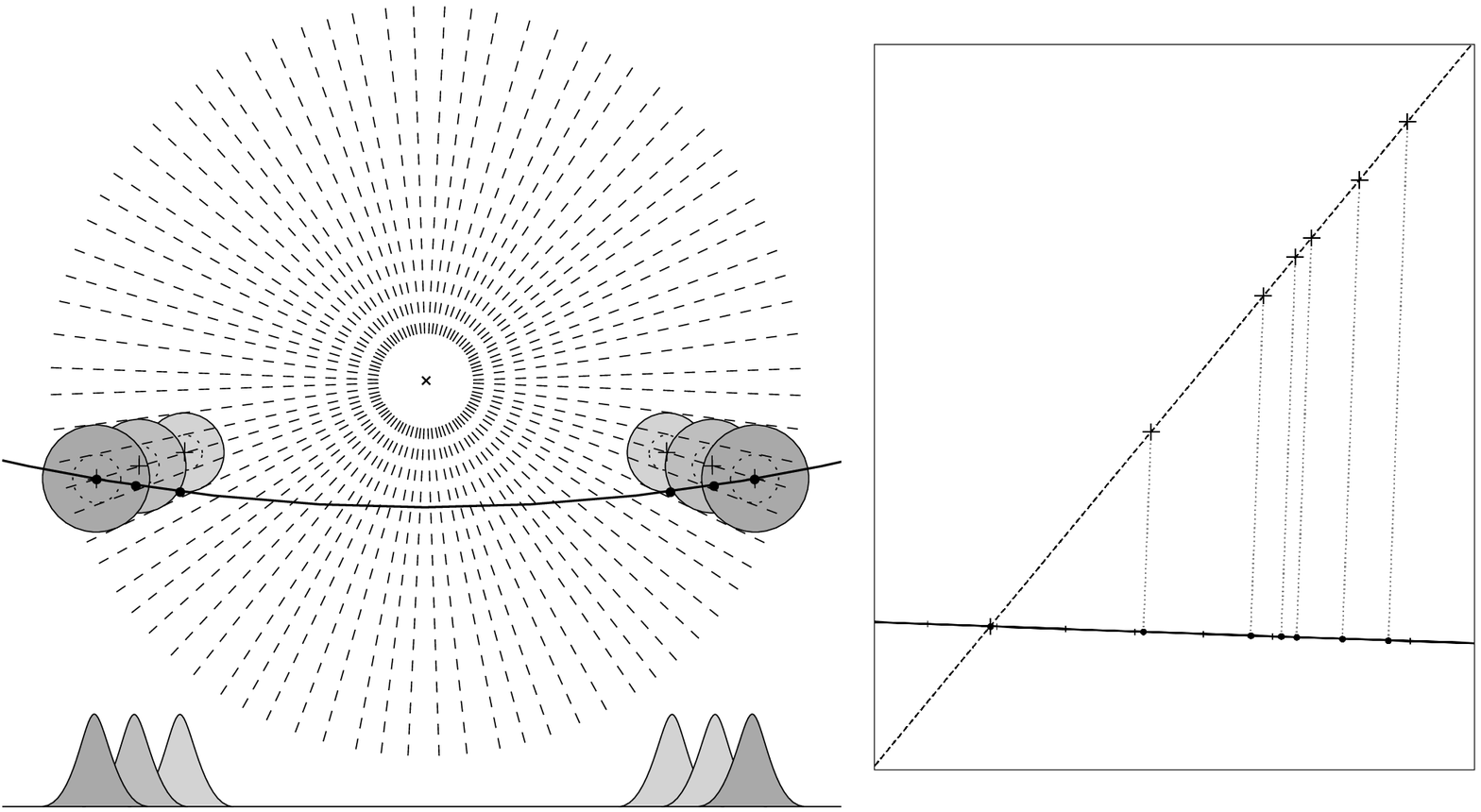}
  \caption{Left: close-up of the same region of the projection down the magnetic axis as in Fig. \ref{fig:hc_projection}, now demonstrating the fan beam emission geometry. The circular contours (shaded regions with dotted inner contours) show the projected emission regions at three frequencies, with darker shading corresponding to lower frequency. The emission region peaks are marked by crosses. The peaks of the observed profiles, depicted below the projection, correspond to the peaks of the regions cut by the line of sight, which, when offset from the actual peak of the emission region, are given by where the contours lie tangent to the line of sight (marked by points). See \citetalias{Dyks2015} for similar diagrams. Right: the same projection down the magnetic axis as in Fig. \ref{fig:hc_projection} with the observed subpulse distribution means and inferred emission region peaks for all seven observing frequencies, now obtained using the fan beam model. The tick marks along the lines of sight are spaced at intervals of $\frac{\pi}{4000}$ radians, going from $\frac{-73}{4000}\pi$ on the right to $\frac{-80}{4000}\pi$ on the left.}
  \label{fig:fan_projection}
\end{figure*}

The mechanism of frequency widening resulting from fan beams is comprehensively discussed in the paper by \citetalias{Dyks2015}. 
Their model results in greater frequency widening being seen for subpulses positioned further away from the fiducial plane; the opposite of the effect seen in the hollow cone model employed above.

The fan beam model of \citetalias{Dyks2015} describes streams of emission diverging away from the magnetic axis, resulting in wedge shaped emission regions on the projection down the magnetic axis of the pulsar. Applying a spectral structure to these streams results in the frequency widening we witness in the data. Considering contours of constant intensity, we can see that a spectral gradient would result in these sets of contours at different frequencies being offset along the stream with respect to each other. These contours could be roughly described as elliptical, offset from each other along the field line that passes through the major axis of each ellipse. The peak of the emission we see would therefore be at the point where the line of sight is tangential to one of these elliptical contours. Since we have no a priori information about the shapes of these elliptical contours, we simplify the model by making the contours circular. We are therefore able to identify the point along the field line corresponding to the intensity peak at a given frequency by considering the projection of the field perpendicular to the magnetic axis, at the emission height. In this projected plane we extend a line perpendicular to the line of sight at the point of the observed peak until it intersects our reference field line $F_{\rm peak}$. This gives us the relationship between $\theta_{\mu}^{j}$ and $\theta_{\rm peak}^{j}$.
The equations describing this process of obtaining $\theta_{\rm peak}^{j}$ from the phase $\phi$ of the observation are given in Appendix \ref{sec:appendix2}.
The extrapolation perpendicular to the line of sight towards the centre of the contours is shown diagrammatically in Fig. \ref{fig:fan_projection}. 

\section{Results}
\label{sec:results}

\subsection{Independence of subpulses}

\begin{figure}
    \includegraphics[width=\columnwidth]{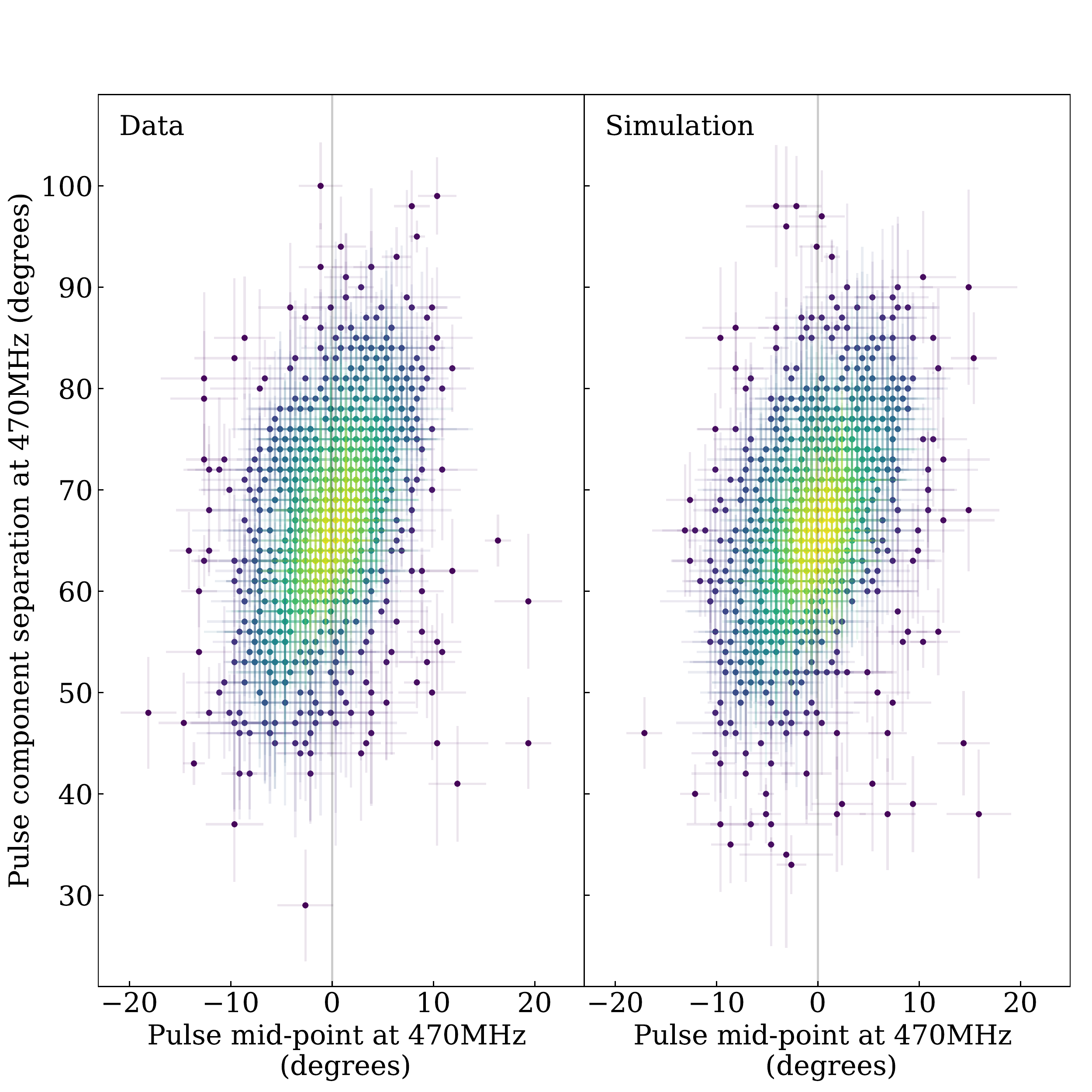}
    \caption{
    Left: scatter plot of subpulse separation against the mid-point between the subpulses for all single pulse profiles in the highest frequency channel 470 MHz. Right: equivalent scatter plot for simulated single pulses, generated by randomly combining the two subpulse distributions
    }
    \label{fig:sepmid}
\end{figure}

It is important to know the relationship between the locations of the two subpulses making up a single pulse. 
If there were a correlation between their positions this would have ramifications both for our understanding of the pulsar emission mechanism and for how our simulations would need to be constructed. 
In Fig. \ref{fig:sepmid} we plot the phase separation between the two subpulses at 470~MHz against the pulse mid-point for each of the single pulses. The mid-point is defined as the position of the point half-way between the two subpulses, such that a mid-point of 0 corresponds to the subpulses being symmetrically positioned about our defined fiducial plane. 
The shading in this figure, and in all of the subsequent scatter plots, indicates the densities of the scatter points, calculated using Gaussian kernel density estimation. This aids in comparison between the distributions of variables for the data and the respective simulations.
Randomly combining the two sets of subpulses and plotting the resulting values for pulse mid-point and separation generates the same distribution as that from the data. The positions of the two subpulses making up a single pulse are therefore independent of each other.

\subsection{Emission region: hollow cone}

\begin{figure*}
    \includegraphics[width=\textwidth]{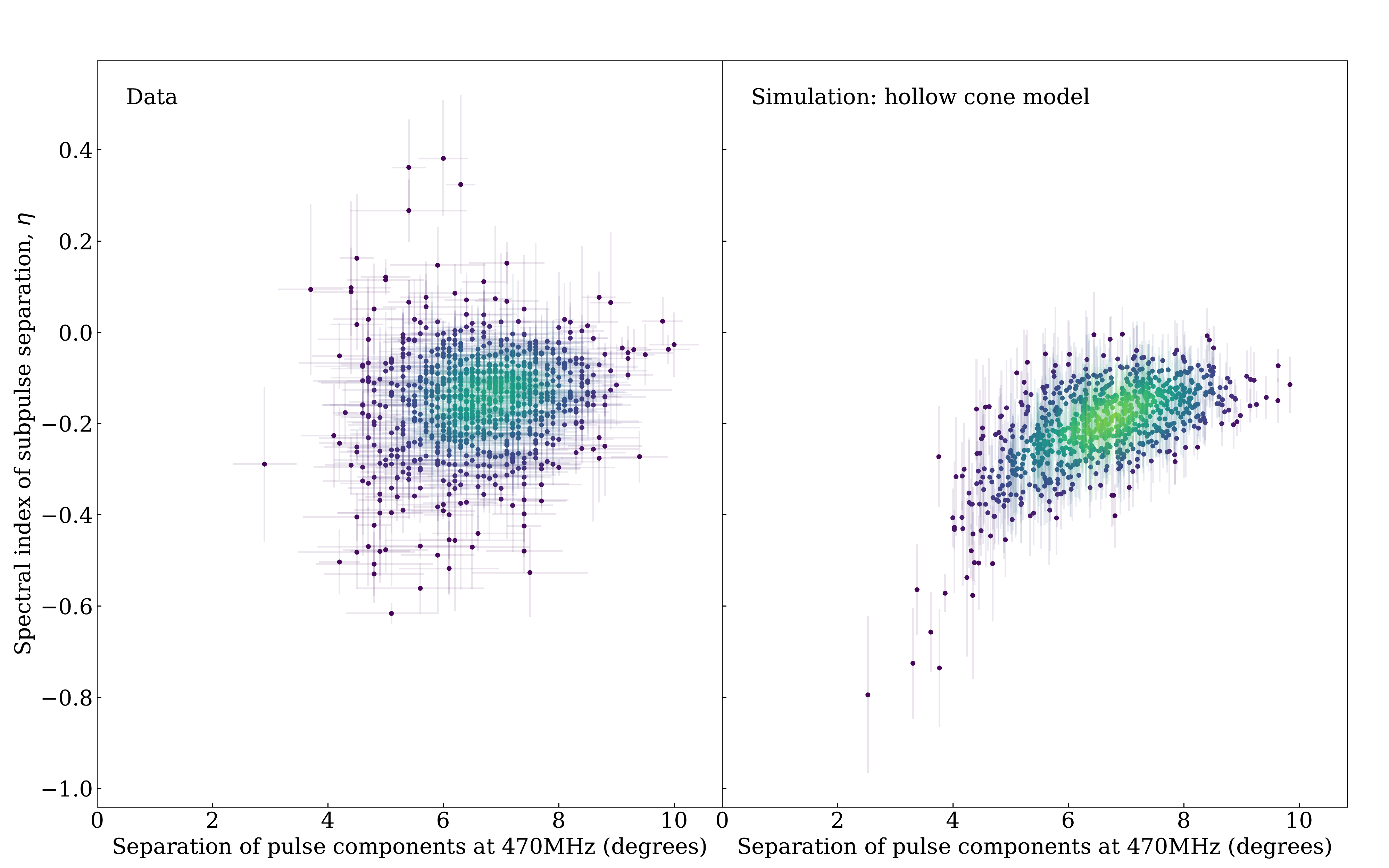}
    \caption{Scatter plots showing the distribution of spectral index of subpulse separation, $\eta$, against pulse separation at the highest frequency channel (470 MHz), for the data (left) and the RFM simulation using the hollow cone emission region (right). The plots are shaded to show the density distribution of points more clearly, with the distribution calculated using Gaussian kernel density estimation. The simulation has a clear slope towards more negative spectral indices at lower separations that is not seen in the data.}
    \label{fig:r_spinsep}
\end{figure*}

Simulating the single pulse frequency evolution expected for the hollow cone model fixes the emission to a ring of field lines with footprint parameters $s^{i}_{L}$ distributed about $s^{p}_{L} = 0.50$ with standard deviations of 0.02 and 0.04 for the two subpulse distributions.
Fig. \ref{fig:r_spinsep} shows the relationship between spectral index of subpulse separation, $\eta$, and pulse peak separation at the highest frequency of the observations (470 MHz) for both the data and the simulation generated with the hollow cone model emission region. Whilst the distributions occupy roughly similar regions of parameter space, there is a clear curved shape in the simulated distribution, such that smaller pulse separations result in steeper spectral indices, in contradiction both with these observations and with those done on multi-component profiles by \cite{Mitra2002}. This is an effect arising from how the line of sight cuts ring-shaped regions of different sizes for this particular geometry. 

As can be seen in Table \ref{tab:musig}, although the mean positions of the subpulse distributions for the hollow cone model are similar to those of the data, the standard deviations of the subpulse distributions do not increase with decreasing frequency as expected, rather they decrease. This too is a geometrical effect of a ring-shaped emission region for this pulsar and is not seen in the data.

\subsection{Emission region: fan beam}
\label{sec:fanresults}

\begin{figure*}
    \includegraphics[width=\textwidth]{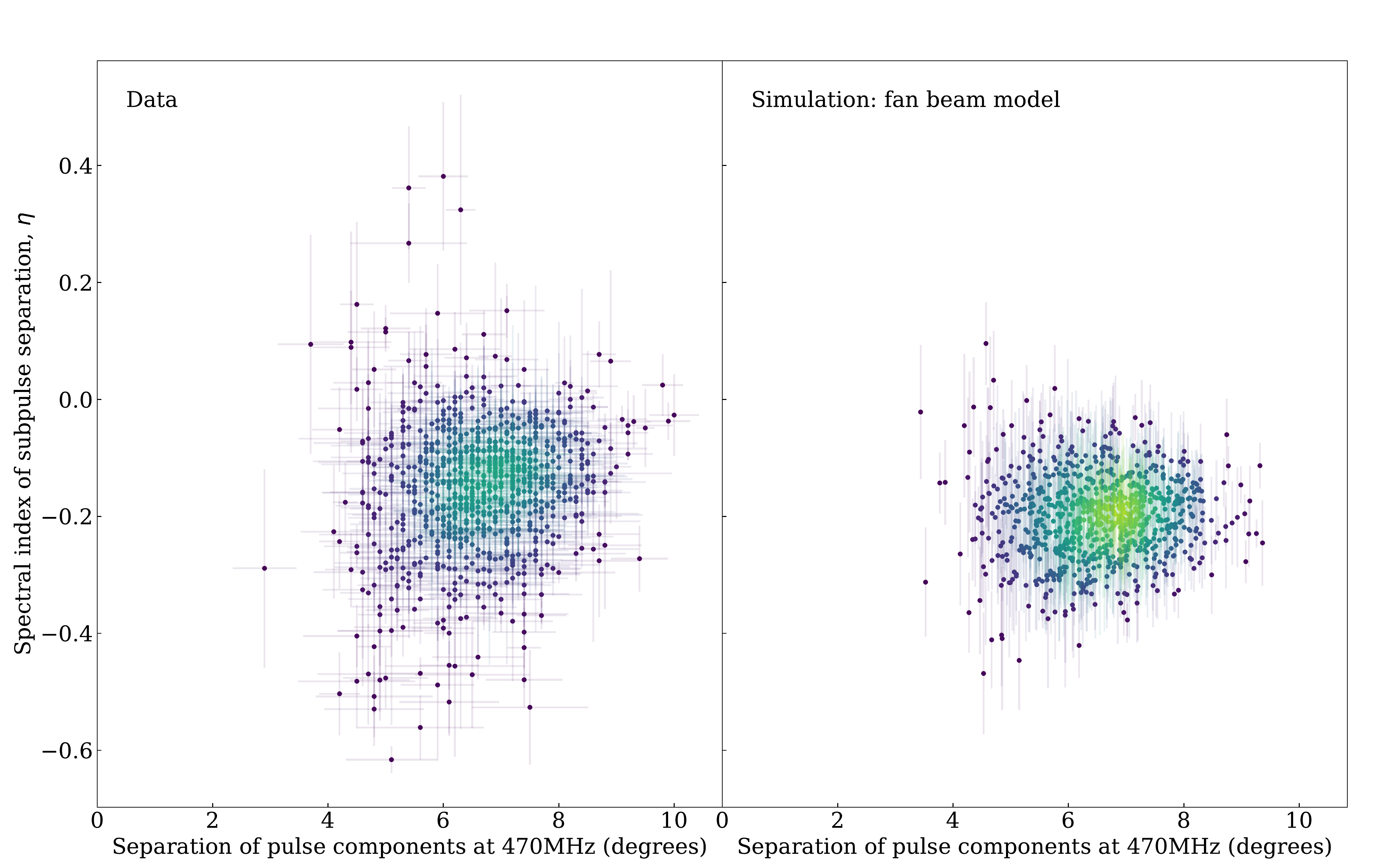}
    \caption{Scatter plots showing the distribution of $\eta$ against pulse separation as for Fig. \ref{fig:r_spinsep}, now with the fan beam model for the emission region.}
    \label{fig:f_spinsep}
\end{figure*}

\begin{figure}
    \includegraphics[width=\columnwidth]{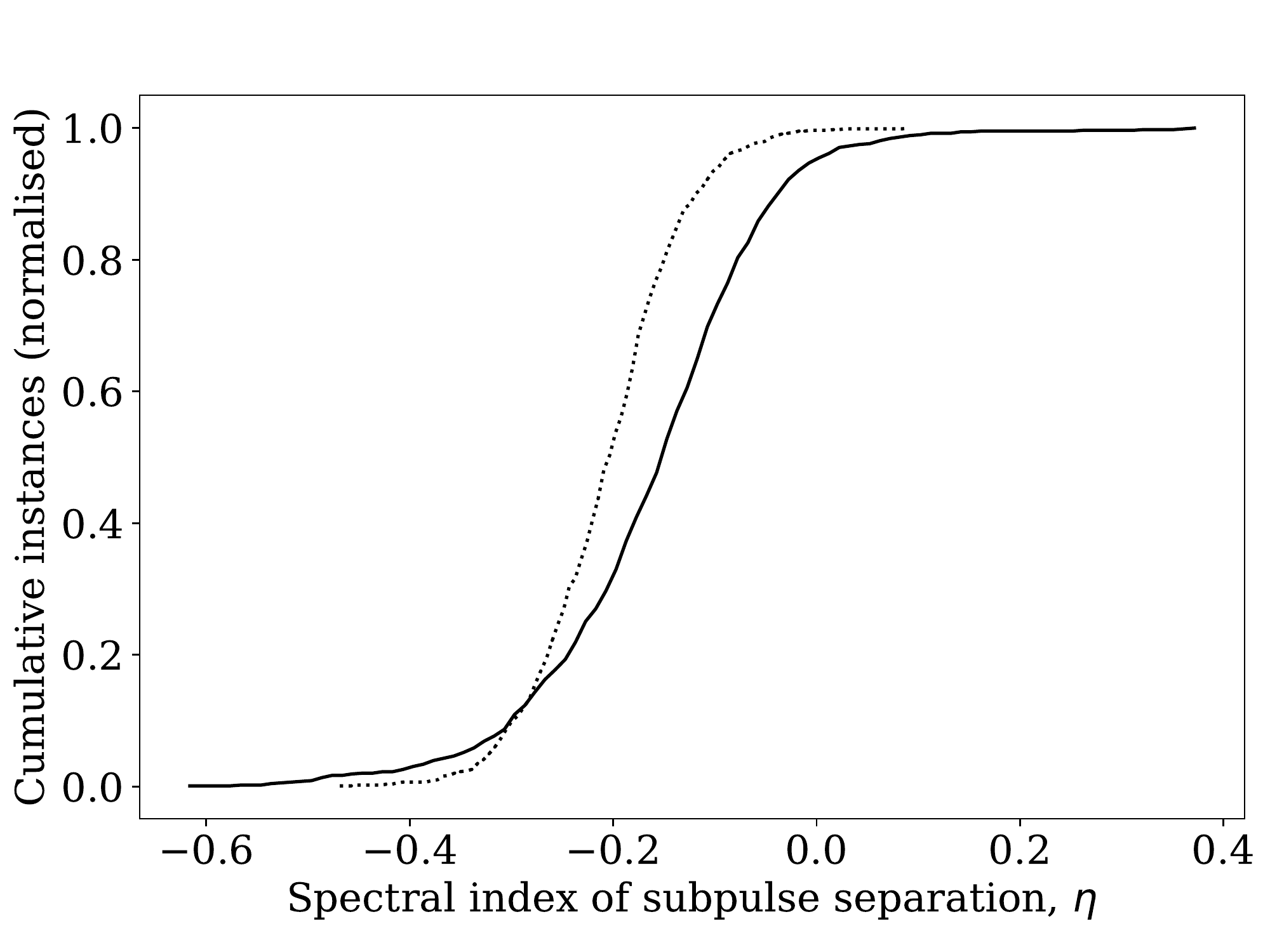}
    \caption{Cumulative distributions of spectral index of subpulse separation---$\eta$---values for the data (solid) and the RFM model with the fan beam emission region (dotted).}
    \label{fig:f_spin}
\end{figure}

For the two fans in the fan beam simulation, the distributions of $s^{i}_{L}$ have mean 0.50 and standard deviations 0.03 and 0.05.
It can clearly be seen in Fig. \ref{fig:f_spinsep} that the relationship between $\eta$ and subpulse separation at 470~MHz for the fan beam model is more similar to the results from the data than those of the hollow cone model. The distributions of subpulse positions also replicate those of the data more closely (see Table \ref{tab:musig}). 

The geometrical construction of the fan beam model produces results that are more compatible with the single pulse frequency evolution. 
However, collapsing Fig. \ref{fig:f_spinsep} along the x-axis to look at the distribution of $\eta$ indicates that the model predicts overly steep frequency widening, as shown in the cumulative plot in Fig. \ref{fig:f_spin}. In particular, the model makes no allowances for $\eta$ at or close to 0, so that a two-sample Kolmogorov-Smirnov test on the two distributions yields a negative result.

Since at different frequencies the peaks of the intensity contours are at different distances from the line of sight, the fan beam model predicts that the intensity of the emission will also vary with frequency as a purely geometrical effect. The further the peak of the emission contours from the line of sight, the lower the intensity observed. The assumption we made in the modelling---that the emission peaks lie on the line of sight at the lowest observed frequency---means that for our simulation the higher frequency emission lies further from the line of sight than the lower frequency emission, so that intensity decreases as frequency increases. We do not have flux calibrated data, but in terms of a qualitative trend this is as expected.
However, this means that the intensity would also decrease for frequencies below our observing band. 
Moving all contour peaks off the line of sight and closer to the magnetic axis would be more correct, but there is no information as to where they should be placed. 
Positioning the contour peaks on the line of sight for the lowest observing frequency therefore gives an upper bound on the resultant emission heights estimated.

\subsection{Orthogonal polarization modes}

\begin{figure*}
    \includegraphics[width=\textwidth]{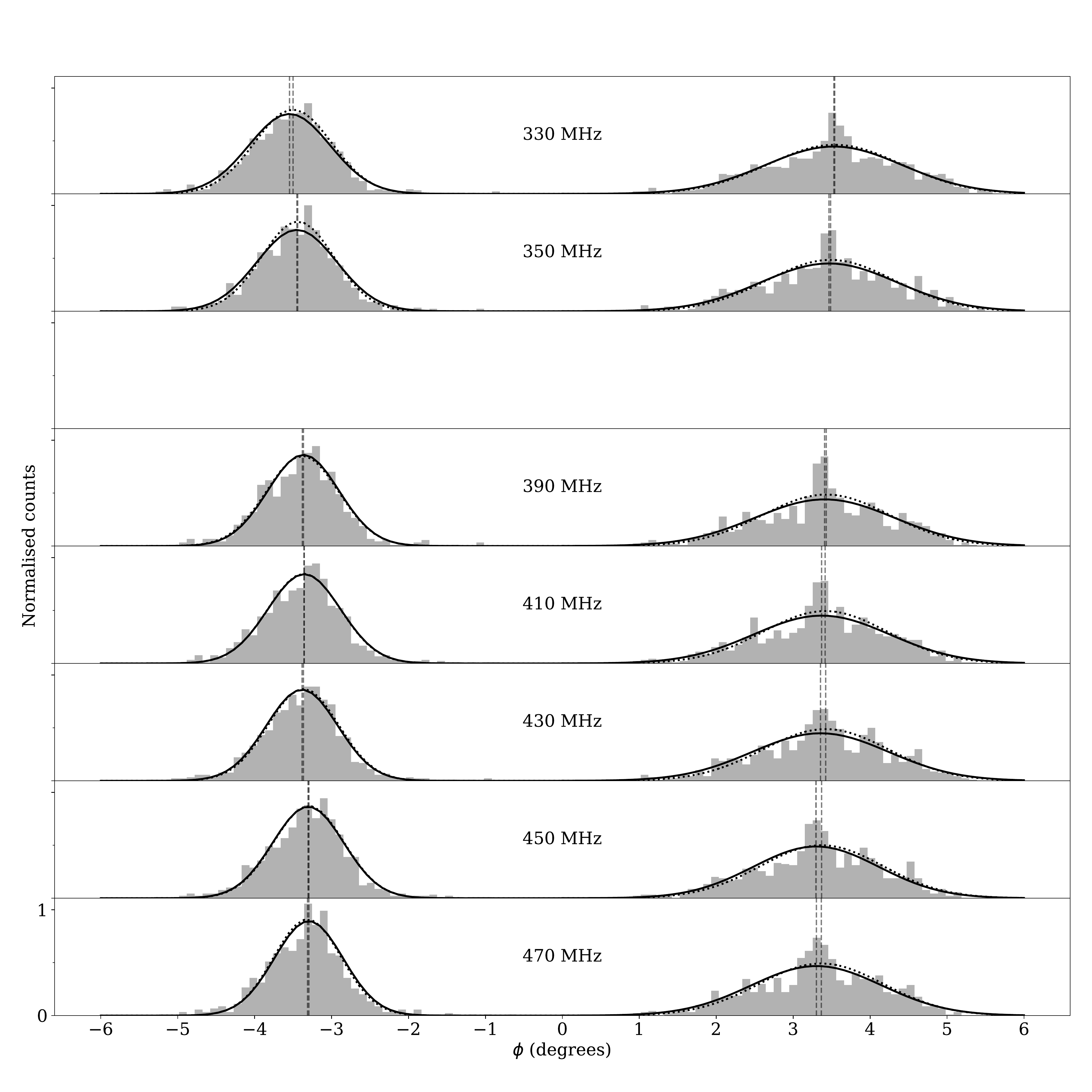}
    \caption{Histograms of subpulse positions generated by the mode divergence simulation with fan beam emission region at all seven frequency bands. Gaussian fits to histograms are over-plotted as lines: the fit to the data as a solid line and the fit to the simulation as a dotted line. See Table \ref{tab:musig} for the means and standard deviations of both sets of Gaussians.}
    \label{fig:m_hist}
\end{figure*}

\begin{table*}
\caption{Table of parameters (mean $\mu$ and standard deviation $\sigma$) of the Gaussian fits to the subpulse distributions for the data and all three simulations.}
\label{tab:musig}
\begin{tabular}{llccccccc}
 \hline
 \multicolumn{9}{l}{\textbf{Subpulse distribution 1 (left)}}\\
 \hline
 \multicolumn{2}{l}{\textbf{Frequency (MHz)}} & 470 & 450 & 430 & 410 & 390 & 350 & 330 \\
 \hline
 Data & $\mu$ ($\degr$)  &  $-$3.30 & $-$3.30 & $-$3.38 & $-$3.35 & $-$3.37 & $-$3.45 & $-$3.55 \\
 
 & $\sigma$ ($\degr$)  & 0.45 & 0.46 & 0.46 & 0.47 & 0.46 & 0.52 & 0.53 \\
 
 Hollow cone & $\mu$ ($\degr$)   &  $-$3.31 & $-$3.30 & $-$3.38 & $-$3.38 & $-$3.39 & $-$3.45 & $-$3.54 \\
 
 & $\sigma$ ($\degr$)   &  0.50 & 0.51 & 0.51 & 0.51 & 0.51 & 0.49 & 0.50 \\
 
 Fan & $\mu$ ($\degr$)  & $-$3.27 & $-$3.26 & $-$3.35 & $-$3.34 & $-$3.38 & $-$3.44 & $-$3.54 \\
 
 & $\sigma$ ($\degr$)   & 0.50 & 0.51 & 0.51 & 0.51 & 0.51 & 0.49 & 0.50 \\
 
 Fan,  & $\mu$ ($\degr$)   & $-$3.31 & $-$3.30 & $-$3.37 & $-$3.36 & $-$3.38 & $-$3.44 & $-$3.50 \\
 
 mode divergence & $\sigma$ ($\degr$)   & 0.44 & 0.46 & 0.46 & 0.47 & 0.47 & 0.47 & 0.50 \\
 \hline
 \multicolumn{9}{l}{\textbf{Subpulse distribution 2 (right)}}\\
 \hline
 \multicolumn{2}{l}{\textbf{Frequency (MHz)}} & 470 & 450 & 430 & 410 & 390 & 350 & 330 \\
 \hline
 Data & $\mu$ ($\degr$)  & 3.30 & 3.30 & 3.35 & 3.37 & 3.41 & 3.46 & 3.53 \\
 
 & $\sigma$ ($\degr$)  & 0.85 & 0.82 & 0.89 & 0.88 & 0.91 & 0.88 & 0.89 \\
 
 Hollow cone & $\mu$ ($\degr$)   & 3.29 & 3.29 & 3.35 & 3.35 & 3.38 & 3.45 & 3.52 \\
 
 & $\sigma$ ($\degr$)   & 0.92 & 0.91 & 0.90 & 0.91 & 0.93 & 0.92 & 0.91 \\
 
 Fan & $\mu$ ($\degr$)   & 3.30 & 3.28 & 3.36 & 3.34 & 3.38 & 3.44 & 3.53 \\
 
 & $\sigma$ ($\degr$)   & 0.83 & 0.85 & 0.85 & 0.86 & 0.87 & 0.89 & 0.90 \\
 
 Fan,  & $\mu$ ($\degr$)   & 3.36 & 3.37 & 3.42 & 3.42 & 3.43 & 3.49 & 3.54 \\
 
 mode divergence & $\sigma$ ($\degr$)   & 0.81 & 0.80 & 0.82 & 0.81 & 0.82 & 0.82 & 0.86 \\
 \hline
\end{tabular}
\end{table*}

\begin{figure*}
    \includegraphics[width=\textwidth]{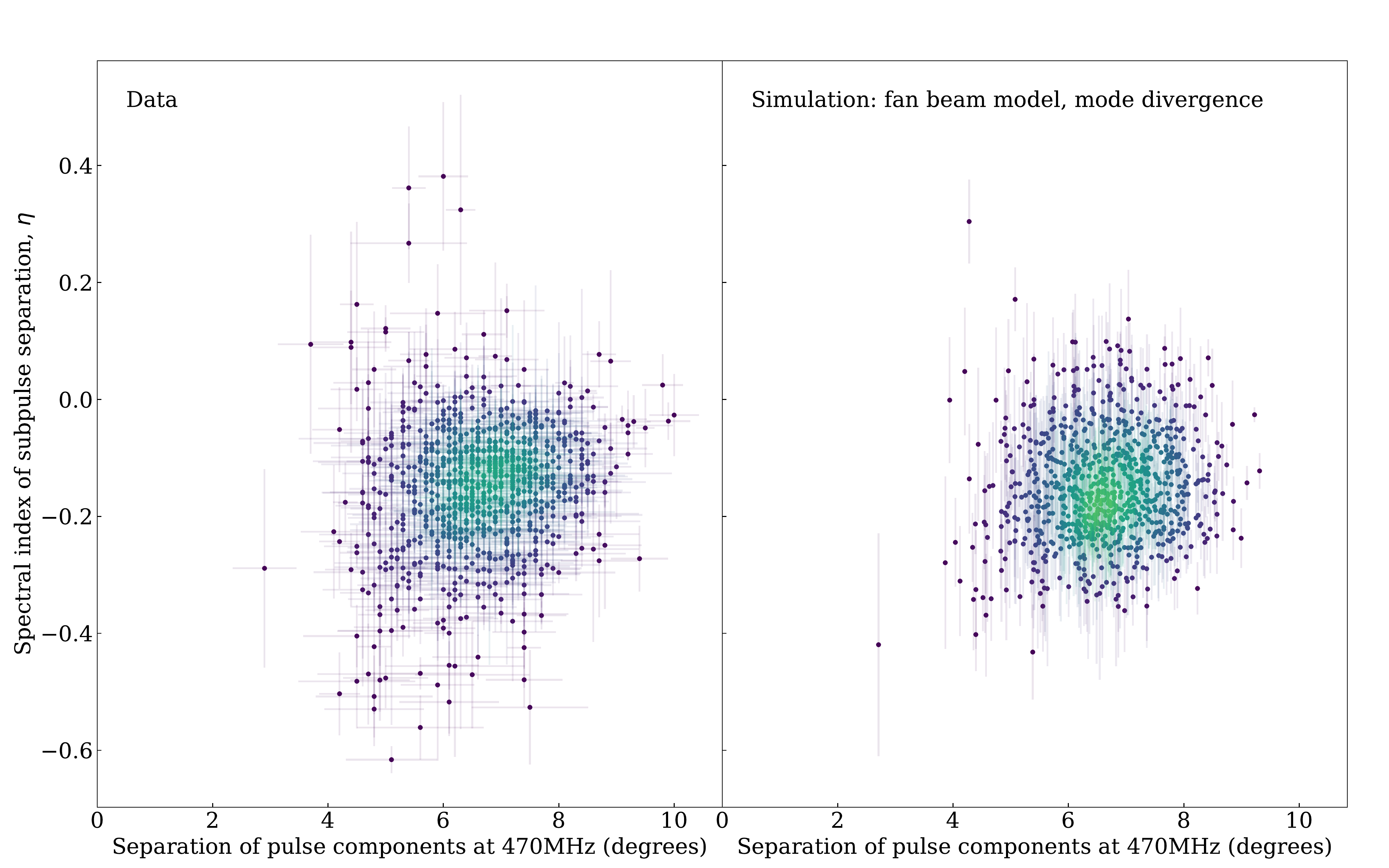}
    \caption{Scatter plots showing the distribution of spectral index of subpulse separation, $\eta$, against pulse separation at the highest frequency channel (470 MHz), for the data (left) and the mode divergence simulation with fan beam emission region (right).}
    \label{fig:m_spinsep}
\end{figure*}

\begin{figure}
    \includegraphics[width=\columnwidth]{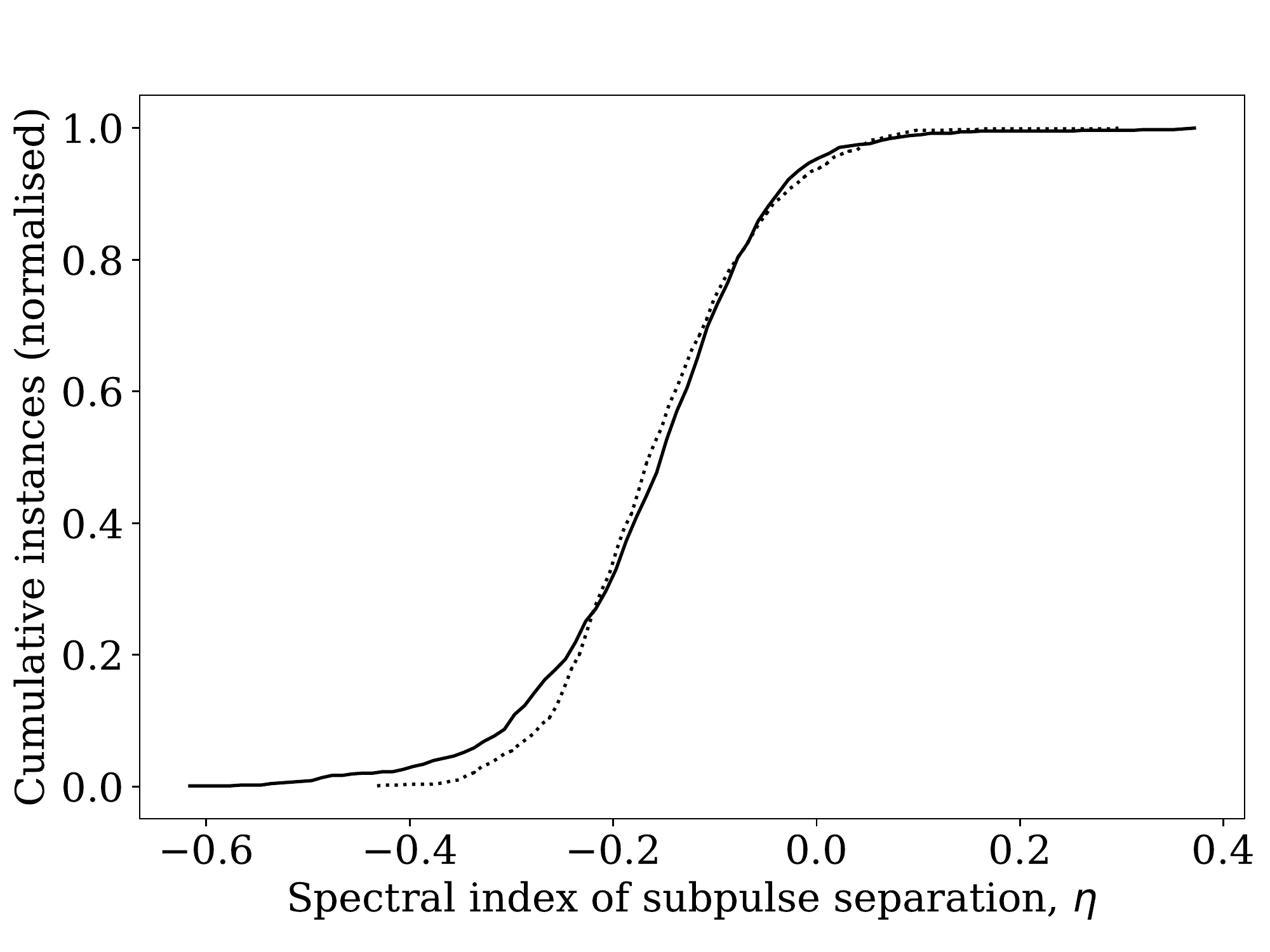}
    \caption{Cumulative distributions of the spectral index $\eta$ for the data (solid) and the mode divergence model with fan beam emission region (dotted).}
    \label{fig:m_spindex}
\end{figure}

The distribution of $\eta$ for the data is skewed towards flatter spectral indices than those predicted by the fan beam model. The theory of OPMs provides a justification for this asymmetry, to account for the frequency-independence of the X mode propagation.
The polarization properties of pulsars, such as jumps of $90\degree$ in the PA and the depolarization of many pulsar profiles at higher frequencies, are commonly explained by the radio emission being composed of two superposed OPMs, as originally suggested by \cite{Manchester1975}. They studied the polarization properties of the single pulses of twelve pulsars, including PSR~J1136+1551, and found that for this pulsar, ``about 25\%'' of the single pulses had a polarization angle orthogonal to that of the other 75\% when averaging over long time intervals, suggesting an approximate 3:1 split in mode dominance for the single pulses of PSR~J1136+1551.

We split the simulated subpulse distributions into X and O modes by assuming that both modes have the same distribution at our highest observing frequency of 470~MHz. For the X mode we therefore randomly draw a fraction of the subpulses from the 470~MHz phase distribution and draw the same distribution at all the other frequencies, since the X mode does not evolve with frequency. What remains is the O mode distribution at each frequency. We then use the positions of the O mode subpulses at the lowest frequency, 330~MHz, to simulate the subpulse phases at all frequencies as described for the fan beam model in sections \ref{sec:method} and \ref{sec:fan}. We then include the X mode subpulses phases at every frequency to obtain the full set of phases for both subpulse distributions.

The simulation requires as an input parameter the value of the height of emission at every frequency. If we use those heights calculated from the average frequency evolution of the O mode distribution, we find that the height range is too broad and results in an overly steep distribution of $\eta$. However, our assumption that the X and O mode subpulses have identical distributions at the highest frequency is likely to result in an overestimation of the emission height range for the O mode pulses alone.
Since we are unable to misalign the distributions of X and O modes at the highest frequency, we instead make use of the heights calculated earlier for the fan beam model without mode divergence. Using these heights and an X mode fraction of 30\% gives the closest replication of the data, which we display here.

Fig. \ref{fig:m_hist} shows the distributions of simulated subpulses from the fan beam model with mode divergence included, with the Gaussian fits to the data distributions overlaid. The subpulse distributions closely follow the shapes of the data distributions, with the left component distribution in particular showing the same asymmetry. The values of the fit parameters, the mean $\mu$ and standard deviation $\sigma$, are given in Table \ref{tab:musig}.
As before, the scatter plot of $\eta$ against separation in Fig. \ref{fig:m_spinsep} is morphologically very similar for the model and the data. It is clear from Fig. \ref{fig:m_spindex} that the distribution of $\eta$ for this model replicates the data much more closely, with a two-sample Kolmogorov-Smirnov statistic of 0.07.

\section{Discussion}
\label{sec:disc}

We have tested RFM in the context of single pulses. Doing so gives us the ability to discriminate between different models of the emission region, within the limitations of the assumptions made.
We have compared the key metrics drawn from the observations with those resulting from simulations of different emission region models: hollow cone and fan beam, with additional frequency-dependent behaviour described by the propagation of OPMs along different paths. 
With respect to our metrics, the fan beam with OPM divergence simulation bears the closest resemblance to the data.
This has implications for our understanding of the pulsar emission mechanism and the inferred heights of emission.

\subsection{Heights of emission}
\label{sec:emheights}

\begin{figure}
    \includegraphics[width=\columnwidth]{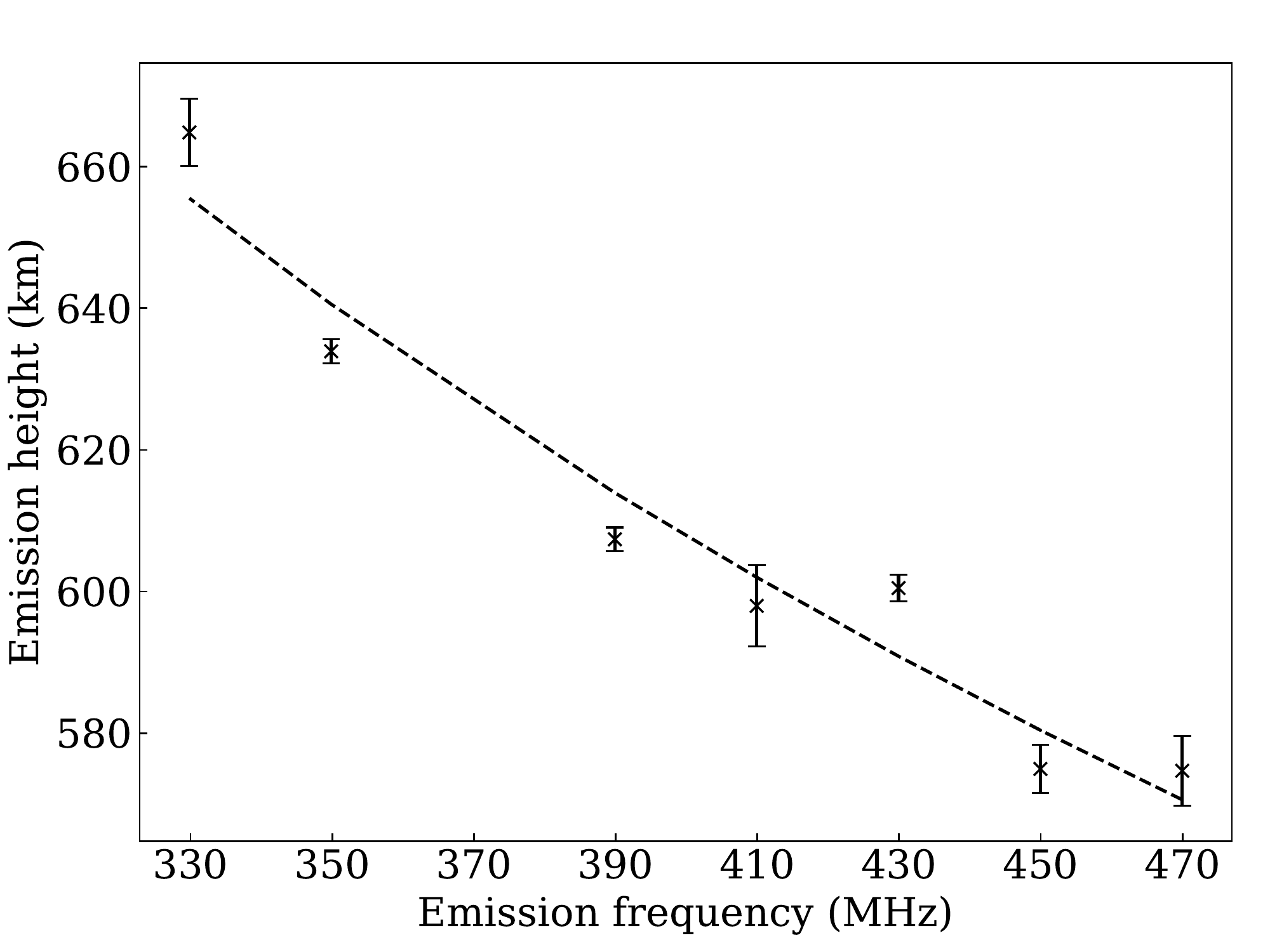}
    \caption{Emission heights calculated using the RFM model with fan beam emission region and $s^{p}_{L}~=~0.5$. The dashed line shows the power law best fit.}
    \label{fig:heights}
\end{figure}

\begin{figure}
  \includegraphics[width=\columnwidth]{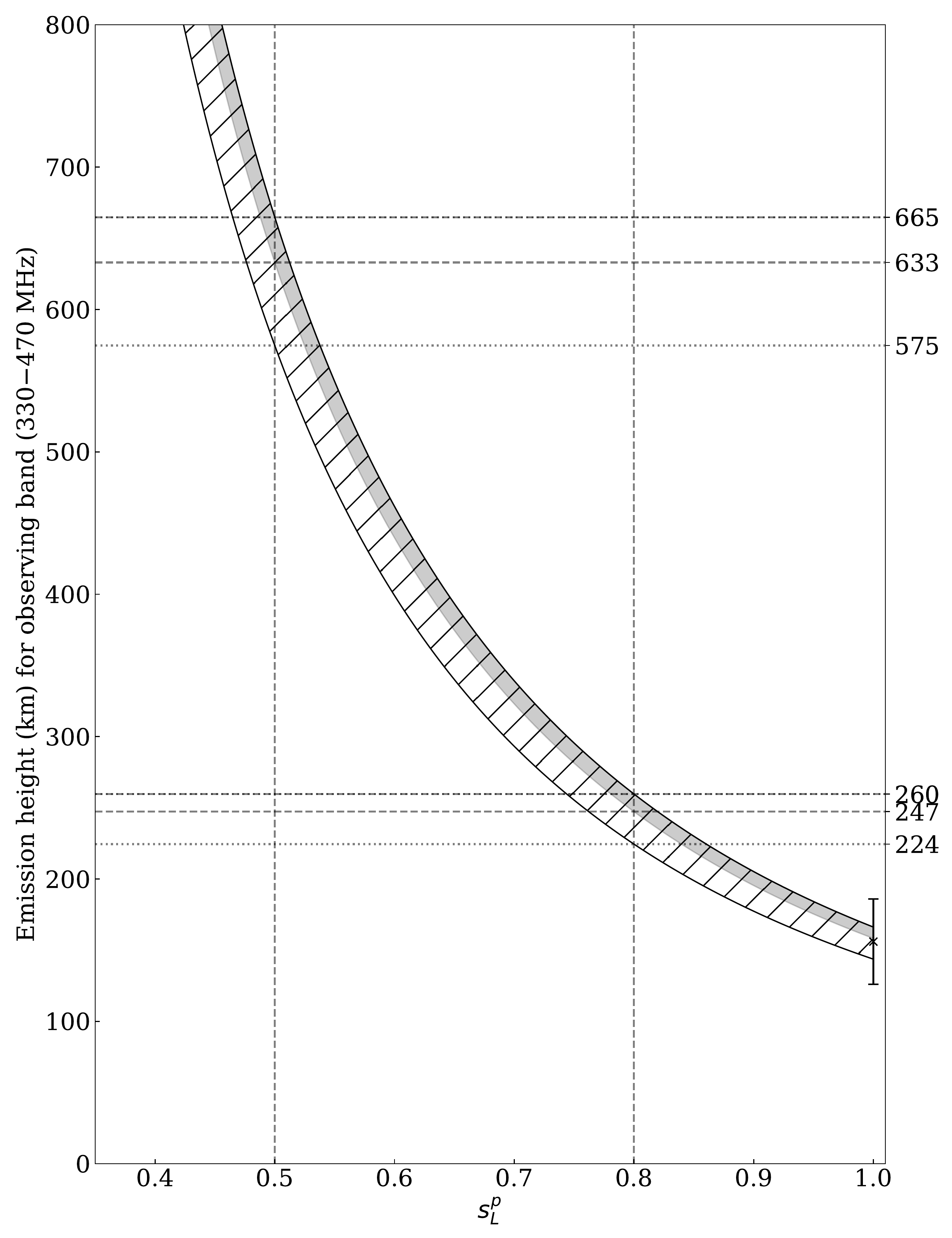}
  \caption{Emission heights over the observing band 330--470~MHz for different values of $s^{p}_{L}$, for both the hollow cone (grey) and fan beam (hatched) models. The emission height calculated by \citetalias{Noutsos2015} at 400~MHz, using the hollow cone model with $s^{p}_{L} = 1$, is overlaid. Dashed horizontal lines mark the highest and lowest heights for the band for each model, corresponding to 330~MHz and 470~MHz respectively, for $s^{p}_{L} = 0.5$ and $s^{p}_{L} = 0.8$ (vertical dashed lines).}
\label{fig:height_footprint}
\end{figure}

Fig. \ref{fig:heights} shows the emission heights generated assuming a fan beam model and $s^{p}_{L} = 0.5$. The best power law fit to the heights, $h$, measured in $km$, against frequency, $f$ (GHz), is given by:
\begin{equation}\label{eq:hVsf}
    h(f)=424~f^{-0.39},
\end{equation}
where the constant of proportionality depends on $s^{p}_{L}$.
These heights are actually an upper bound, caused by fixing the peak of the emission region for the lowest observed frequency to the line of sight as discussed in section \ref{sec:fanresults}. Shifting the peaks closer to the magnetic axis would reduce the emission heights but would not substantially alter the range over which these heights are spread.
The fan beam model results in a larger spread of heights across the observing band than the hollow cone model. This can be readily understood by considering the geometry on the 2D projection of the beam: see Fig. \ref{fig:hc_projection} and Fig. \ref{fig:fan_projection}. The observed peaks of emission from the fan beam model are related to points further along the reference field line, and therefore at a greater height difference, than the equivalent for the hollow cone model.

For the single pulse analysis we have fixed $s^{p}_{L}$ to 0.5, however we can obtain qualitatively similar metrics using a different value, which shifts the emission heights and height ranges. In other words, our results do not constrain $s^{p}_{L}$, and allow us the freedom to change its value given additional information. Fig. \ref{fig:height_footprint} shows how the heights and height ranges compare for the two models, and how these values change with $s^{p}_{L}$. Previous work modelling the emission heights of the integrated profile of PSR~J1136+1551 has used the hollow cone model with the emission fixed to the last open field lines so that $s^{p}_{L} = 1$. We show the emission height that \citetalias{Noutsos2015} calculated for a frequency of 400~MHz, using the component peak separation, in Fig. \ref{fig:height_footprint}, which is in agreement with our measurements for $s^{p}_{L} = 1$. 

\subsection{Aberration and retardation}
\label{sec:a_r}

Relativistic aberration and retardation (A/R) effects are responsible for a phase lag between the intensity and PA profiles at a given frequency, such that $\Delta \phi = 4\frac{r}{R_{LC}}$ (see \citealt{Blaskiewicz1991,Johnston2006}). \citetalias{Noutsos2015} measure this phase lag for PSR~J1136+1551 at 150~MHz, defining the fiducial point of the intensity profile as the mid-point between the two component peaks and that of the PA as the point where it has the steepest gradient. This gives an emission height of $349 + 158 - 150$~km. To be consistent with this measurement requires a change in analysis from  $s^{p}_{L} = 0.5$ to $s^{p}_{L} = 0.8 + 0.2 - 0.1$. The height shift that this entails is shown in Fig. \ref{fig:height_footprint}. In the absence of polarimetric data, and given that we have dedispersed and aligned our data such that the means of the subpulse distributions evolve symmetrically with respect to the magnetic axis, we cannot directly measure A/R effects.

\cite{Kramer1997} measured an upper limit on the pulse time-of-arrival difference of 1640~$\mu$s between 1.41 and 32~GHz, which they translated to a height range of 310~km using $\alpha = 147.0\degr$ (for $\alpha = 51.3\degr$ this translates to 276~km). Extrapolation of our heights for the fan beam model to their frequency band using equation \ref{eq:hVsf}, results in height ranges below their upper limit for all values of  $s^{p}_{L} \geq  0.5$.

\citetalias{Hassall2011} measured no time delay attributable to A/R across the lower frequency range of 40-180~MHz, which translates to an upper bound of the height range of 59~km. This is very low compared to our extrapolated height range. In principle, we might expect measurable A/R in their data. The low heights they postulate may be attributed to their technique for determination of the fiducial point and dedispersion, absorbing the effect.

All considerations of the effects of A/R so far have assumed a strict one-to-one mapping of radius and frequency. However it is already clear from our simulations that the single pulse behaviour of PSR~J1136+1551 is better explained by there being two emission paths which we associate with OPMs. A contribution from the frequency-independent X mode would reduce the magnitude of A/R measured in comparison to that which would otherwise be expected. \citetalias{Noutsos2015} refute the possibility that the frequency widening of PSR~J1136+1551 may be caused by diverging OPMs, suggesting that the lack of jumps in the PA profile at frequencies below 1400 MHz is possible evidence against this being the widening mechanism.  There are, however, emission geometries and X and O mode distributions that could result in the amplitude of one being everywhere smaller than that of the other along the line of sight. This would mean a net widening could be seen without the need for any mode jumps.
Furthermore, if emission at a given frequency comes not from a single height but from some range, the frequency dependence of a timing A/R effect would be far less clear than otherwise expected. Polarization data can be used to clarify these issues.

\subsection{Emission region shape}

The results of our simulation favour the fan beam model, a description of the emission region that is independently favoured by the work of \cite{Wang2014}, in their investigations of precessing pulsars. 
Unlike the hollow cone model, it does not provide a clear explanation for the symmetry of the two emission regions about the fiducial plane. However, the symmetrical frequency evolution, which implies symmetrical positioning of the emitting regions, is not directly inferred from the data. It is a result of our choice of DM, the reasoning for which has been discussed in section \ref{sec:fid}. There is therefore some flexibility: the exact positioning and frequency evolution of the subpulse distributions do not have to be perfectly symmetrical in the simulation. Indeed, given that the two distributions have different shapes, it might be surprising if they were. 

We have addressed above the key implications of this model on the emission heights generated, but only in the case of a one-to-one radius-to-frequency mapping. It seems likely that region within which a given frequency is produced could extend over at least a small range of heights.
The model of \citetalias{Gangadhara2004}, which relates frequency to field line curvature instead of height, presents an alternative avenue of consideration. In such a picture the observable radiation must originate from a range of heights, as the line of sight crosses regions at different perpendicular distances from the magnetic axis. A simulation investigating these considerations would more accurately replicate the fan beam model of \citetalias{Dyks2015}, for which each frequency is emitted by an extended region along the beam.

\section{Conclusions}
\label{sec:conc}

Use of the single pulse frequency evolution of PSR~J1136+1551 has enabled us to make direct comparisons of two different models of the emission region within the magnetosphere. The advantages of the hollow cone model can be readily seen: it is simple, symmetrical and generates heights that are both comparable with previous work and do not conflict with our lack of A/R measurement. This however is insufficient to justify its use. The data alignment across frequency is dependent on the choice of DM. In addition, its subpulse distributions have different widths: the symmetry of the data may be imposed rather than inherent. Our metrics describing the single pulse evolution cannot be explained by the hollow cone model, which incorrectly predicts a shrinking of the subpulse distributions with decreasing frequency and a geometry-based relationship between spectral index of separation, $\eta$, and separation of subpulse components at a given frequency (Fig. \ref{fig:r_spinsep}).

By contrast, the fan beam model does not have symmetry requirements and the subpulse evolution it predicts is much closer to than seen in the data. This model is however limited by the constraint of a one-to-one mapping of frequency and height. It also only generates an upper bound on emission heights, since lack of other knowledge necessitates our positioning the lowest frequency contours so that they lie exactly on the line of sight.

The distribution of spectral indices generated by the fan beam model does not replicate the subset of single pulses in the data that have $\eta$ close to zero. This can however be explained by the differing frequency behaviour of OPMs. Our implementation of OPM behaviour is subject to the same limitations as those of the fan beam model, plus the added limitation of having to use the subpulse distribution at the highest observed frequency as our fixed point where the two modes converge in emission height. Despite this, the fan beam model with mode divergence results in the best replication of the single pulse evolution observed in the data.

It is striking that the frequency evolution of PSR~J1136+1551 can be successfully described by such a simple model of the emission region, and that it is possible to make clear distinctions between the effects of different emission region models. Statistical analysis of single pulses provides a means of probing the emission region of pulsars in a way that is otherwise impossible. We intend to apply this technique to more pulsars with similar emission characteristics, and incorporate polarimetric information in future work.

\section*{Acknowledgements}

We wish to thank Dipanjan Mitra for useful discussions and help in performing the observations.
LO acknowledges funding from the Science and Technology Facilities Council (STFC) Grant Code ST/R505006/1.
We thank the staff of the GMRT that made these observations possible. GMRT is run by the National Centre for Radio Astrophysics of the Tata Institute of Fundamental Research.

\bibliographystyle{mnras}
\bibliography{bibliography}

\appendix

\section{Field line constant}
\label{sec:appendix1}

The footprint parameters associated with the emitting field line and the last open field line are related to the height of the neutron star surface $r_{s}$ as follows:
\begin{equation}
    r_{s} = K^{i}\sin^{2}\left(s^{i}/r_{s}\right) = R_{LC}\sin^{2}\left(s_{L}/r_{s}\right).
\end{equation}
Rearranging these equations we can obtain the individual footprint parameters and the footprint parameter ratio $s^{i}_{L}$:
\begin{equation}
    s^{i}_{L} = \frac{s^{i}}{s_{L}} = \frac{\arcsin{\left(\sqrt{r_{s}/K^{i}}\right)}}{\arcsin{\left(\sqrt{r_{s}/R_{LC}}\right)}}.
\end{equation}
We obtain the field line constant $K^{i}$ of the emitting field line by further rearrangement:
\begin{equation}
    K^{i} = \frac{r_{s}}{\sin^{2}\left(s^{i}_{L}\arcsin{\sqrt{r_{s}/R_{LC}}}\right)}.
\end{equation}

\section{Beam half-opening angle for the fan beam model}
\label{sec:appendix2}

Using the spherical law of cosines for angles ($A, B, C$) and sides ($a, b, c$) of a spherical triangle, 
$\cos c=\cos a\cos b+\sin a\sin b\cos C$, we can derive the beam half-opening angle $\rho$ for emission that points along the line of sight in terms of the rotational phase $\phi$, the angle between the magnetic and rotation axes $\alpha$ and the angle between the magnetic axis and the vector pointing along the line of sight at the point of emission, $\beta$:
\begin{equation}
    \cos(\rho) = \cos(\alpha+\beta)\cos(\alpha) + \sin(\alpha+\beta)\sin(\alpha)\cos(\phi).
\end{equation}
We can use the same law to find the angle of intersection of the fiducial plane and the plane containing the magnetic axis and the vector pointing towards the line of sight, as a function of the geometry and the beam half-opening angle $\rho$ at each rotation phase:
\begin{equation}
    \cos(\gamma) = \frac{\cos(\alpha+\beta) - \cos(\alpha)\cos(\rho)}{\sin(\alpha)\sin(\rho)}.
\end{equation}
We can therefore describe the line of sight in the x-y plane projected perpendicular to the magnetic axis as follows:
\begin{equation}
    x_{L} = \rho\sin(\gamma),
\end{equation}
\begin{equation}
    y_{L} = \rho\cos(\gamma).
\end{equation}
Equations describing the tangent to the line of sight at point $(x_{L}, y_{L})$ are obtained through differentiation and the chain rule: 
\begin{equation}
\begin{aligned}
\frac{\partial y_{L}}{\partial\rho} ={} &  \frac{\cos(\alpha+\beta)}{\sin(\alpha)\sin(\rho)} \left(1 - \frac{\rho}{\tan(\rho)}\right) \\
                                        &  + \frac{\cos(\alpha)}{\sin(\alpha)}\left(\rho\left(1 + \frac{1}{\tan^{2}(\rho)}\right) - \frac{1}{\tan(\rho)}\right),
\end{aligned}
\end{equation}
\begin{equation}
\begin{aligned}
\frac{\partial x_{L}}{\partial\rho} ={} & \sqrt{1 - \left(\frac{\cos(\alpha+\beta) - \cos(\alpha)\cos(x_{L})}{\sin(\alpha)\sin(\rho)}\right)^{2}   } \\
                                        & +       \frac{\rho F}{          \sqrt{1 -      \left(\frac{\cos(\alpha+\beta) - \cos(\alpha)\cos(\rho)}{\sin(\alpha)\sin(\rho)}\right)^{2}}},
\end{aligned}
\end{equation}
where F is given by:
\begin{equation}
\begin{aligned}
F ={} & \frac{(\cos(\alpha+\beta) - \cos(\alpha)\cos(\rho))^{2}}{\sin^{2}(\alpha)\sin^{2}(\rho)\tan(\rho)} \\
    & - \frac{\cos(\alpha)(\cos(\alpha+\beta) - \cos(\alpha)\cos(\rho))}{\sin^{2}(\alpha)\sin(\rho)},
\end{aligned}
\end{equation}
\begin{equation}
    \frac{\partial y_{L}}{\partial x_{L}} = \left. \frac{\partial y_{L}}{\partial \rho} \middle/ \frac{\partial \rho}{\partial x_{L}} \right..
\end{equation}
Given an observation at phase $\phi^{i}$ and a reference observation at $\phi^{0}$ we obtain their positions on the line of sight, $(x_{L}^{i}, y_{L}^{i})$ and $(x_{L}^{0}, y_{L}^{0})$ as above.
We assume that the peak of the emission region for the reference observation, $(x_{\rm peak}^{0}, y_{\rm peak}^{0})$, lies at $(x_{L}^{0}, y_{L}^{0})$. We further assume that the peak of the emission region for our observation $\phi^{i}$ lies further along the same field line. We deduce that for a circular emission region, the observation at the line of sight, $(x_{L}^{i}, y_{L}^{i})$, is connected to this emission region peak by a perpendicular bisector to the line of sight at this point. We therefore locate the emission region peak $(x_{\rm peak}^{i}, y_{\rm peak}^{i})$ as being at the intersection of this perpendicular bisector and the field line.
\begin{equation}
    x_{\rm peak}^{i} = \frac{y_{L}^{0} - g_{L}x_{L}^{0}}{g_{f} - g_{L}},
\end{equation}
\begin{equation}
    y_{\rm peak}^{i} = g_{f}x_{\rm peak}^{i},
\end{equation}
where $g_{f}$ is the gradient of the field line and $g_{L}$ is the gradient of the perpendicular bisector to the line of sight, with their values given as follows:
\begin{equation}
    g_{f} = \frac{y_{L}^{0}}{x_{L}^{0}},
\end{equation}
\begin{equation}
    g_{L} = \left.   -1    \middle/   \frac{\partial y_{L}^{i}}{\partial x_{L}^{i}} \right..
\end{equation}
We can now convert back from the projected x-y plane to the beam half-opening angle of the emission region peak for all of our observations:
\begin{equation}
    \rho_{\rm peak} = \sqrt{x_{\rm peak}^{2} + y_{\rm peak}^{2}}
\end{equation}
and half-opening angle $\rho_{\rm peak}$ is converted to spherical polar angle $\theta_{\rm peak}$ using equation \ref{eq:theta}.

\bsp	
\label{lastpage}

\end{document}